\documentclass{ametsocV6.1}

\title{Energy transfers in surface wave-averaged equations}

\authors{Lars Czeschel\aff{a}\correspondingauthor{Lars Czeschel, lczeschel@uni-hamburg.de}
Carsten Eden\aff{a}}

\affiliation{\aff{a}{University of Hamburg}}

\abstract{Ocean surface gravity waves play an important role for the air-sea momentum fluxes and the upper ocean mixing, and knowledge of the sea state leads 
in general circulation models
to improved estimates of the ocean energy budget and allows to incorporate 
surface wave impacts, such as Langmuir turbulence. 
However, 
including the Stokes drift ${\mathbf u}^S$, in phase-averaged equations for the Eulerian mean motion leads to  an Eulerian energy budget which is physically difficult to interpret. In this note, we show that a Lagrangian energy budget allows for a closed energy budget, in which all terms connecting the different energy compartments correspond to well known energy transfer terms. 
We show that the so-called Coriolis-Stokes force does not lead to an energy transfer between surface gravity waves and oceanic mean motions as previously suggested.
Instead, the Coriolis-Stokes force transfers energy between the Eulerian mean kinetic energy,  $MKE_E = \frac{1}{2} \overline{{\mathbf u}}\cdot  \overline{{\mathbf u}}$, and a mean energy compartment which is the product of  the mean Eulerian velocity $\overline{{\mathbf u}}$ and the mean Stokes drift $\overline{{\mathbf u}^S}$, $MKE_{ES} =  \overline{{\mathbf u}}\cdot \overline{{\mathbf u}^S}$. Both energy forms are a result of the unnatural split-up of the Lagrangian velocity into Eulerian velocity and the Stokes drift. In an energy budget for the Lagrangian  mean kinetic energy, the work done by the Coriolis-Stokes force does not contribute, and should be used to estimate the kinetic energy balance in the wave‐affected surface mixed  layer. The Lagrangian energy budget is used to discuss an energetically consistent framework which can be used to couple a general circulation ocean model to a surface wave model.}

\begin{document}

\maketitle

\section{Intoduction}

In a non-rotating frame, \cite{Stokes1847} established that surface gravity waves induce a mean flow in the direction of wave propagation, known as the Stokes drift ${\mathbf u^S}$. 
\cite{Craik76} were able to incorporate ${\mathbf u^S} $ in wave averaged Boussinesq momentum equations for the Eulerian velocity ${\mathbf u} $.  Including the Coriolis force, the equations are given by (e.g. \cite{Huang1979}, \cite{Leibovich1980}):
\begin{equation}\label{eq1}
\partial_t {\mathbf u} + ({\mathbf u} \cdot \nabla) {\mathbf u}   =  - {\mathbf f} \times  {\mathbf u} - \underbrace{{\mathbf f} \times  {\mathbf u^S}}_{F_{CS}} + b {\mathbf z} - \nabla p^*  +  \underbrace{{\mathbf u^S}\times {\boldsymbol \omega}}_{F_{V}}  + {\mathbf D_u} 
\end{equation}
Here, $b=-g \rho/ \rho_0$ denotes buoyancy, ${\boldsymbol \omega} = \nabla \times {\mathbf u}$ is the Eulerian vorticity, $ {\mathbf f}$ is the Coriolis parameter, and ${\mathbf D_u}$ indicates dissipation of Eulerian momentum. Surface wave impacts enter the equation through ${\mathbf u^S}$ and are given by the  Coriolis-Stokes force $F_{CS}$,  the vortex force $ F_{V}$, and the modified pressure  $p^* = p /  \rho_0 + \frac{1}{2} [({\mathbf u} - {\mathbf u^S})^2 - {\mathbf u}^2]$. The equations are usually referred to as 'Craik-Leibovich equations' and are widely used to study the impact of surface waves on the oceanic surface mixed layer and Ekman-spiral solution (e.g. \cite{Skyllingstad1995}, \cite{McWilliams1997},  \cite{Polton2005}).

The Coriolis Stokes force $F_{CS}$ originates from the wave-induced Reynolds stresses, and leads in an inviscid ocean to an Eulerian flow which is exactly opposite to the Stokes drift, resulting in a vanishing Lagrangian mean flow \cite[]{hasselmann1970wave}.  This result is in agreement with previous findings, that in a rotating ocean, a Lagrangian mass transport cannot arise from a steady wave field (\cite{Ursell1950},  \cite{Pollard1970}). \cite{hasselmann1970wave} further established that the Coriolis Stokes force leads to surface wave driven inertial oscillations. 
The vortex force $ F_{V}$ causes vorticity to  tilt in the direction of the Stokes drift. The result are coherent vortices known as  Langmuir circulation  \cite[]{Craik76}. The associated Langmuir turbulence is often a dominant source for turbulent motions and mixing in the oceanic mixed layer \cite[]{belcher2012global}.
Note, that the term 'Langmuir turbulence' sometimes include the shear driven turbulence of the Eulerian return flow, called anti-Stokes flow.

Global and regional climate models do not resolve Langmuir turbulence, however  numerous parameterizations exist and have been tested in such models (e.g. \cite{Fan2014}, \cite{Ali2019}). The parameterizations usually rely on the knowledge of the Stokes drift. Under the assumption of a fully developed sea, the Stokes drift can be approximated by using the local wind. However, such a sea state in equilibrium seems to be a rather poor assumption \cite[]{hanley2010}. Another possibility is to use a third generation surface wave model like WAM \cite[]{komen1996} or \cite{wavewatch2016}. Several studies have coupled such a surface wave model to regional or global climate models 
by using some form of the Craik-Leibovich equation (e.g. \cite{Breivik2015}, \cite{Li2016}, \cite{Sun2022}) or  phase-averaged equations of even higher order in vertical shear (\cite{Couvelard2020}). 

The Stokes drift obtained from surface wave models has also been utilized to estimate the energy input into the ocean.
If the Craik-Leibovich equation (Eq.~(\ref{eq1})) is used to form an energy equation for the Eulerian mean kinetic energy,  $MKE_E = \frac{1}{2} \overline{{\mathbf u}}\cdot\overline{{\mathbf u}}$, a transfer term appears of the form $\partial_t MKE_E = - \overline{{\mathbf u}} \cdot ({\mathbf f} \times \overline{{\mathbf u^S}}) $. The overbar denotes an adequate averaging. The Coriolis force, despite being a fictitious force, contributes then in this energy equation to the energy budget which is difficult to understand. The above transfer term is considered to be an energy transfer from the surface waves to the Eulerian kinetic energy (e.g. \cite{Suzuki2016}, hereafter SFK16) and has been used to calculate the energy input into the mixed layer (\cite{Liu2009wind}, \cite{Sayol2016}, \cite{Zhang2019}, among others).
On the global scale \cite{Liu2009wind} estimated an energy input of $0.29$ Tw in the Ekman layer through the work done by the Coriolis-Stokes force, which is a significant share of other
important transfer rates in the global ocean energy cycle.
However, at least to our knowledge, no energy equation for surface waves was derived which shows the corresponding transfer term of opposite sign.
\cite{Brostrom2014} and  \cite{Weber2015} discussed some inconsistencies in the energy budget related to the Coriolis-Stokes force. They both conclude that the 
Coriolis-Stokes force plays no role in the energy budget, if the  budget is integrated vertically to the moving material surface, i.e. in a Lagrangian framework.

If Eq.~(\ref{eq1})  is used to form an energy equation for the Eulerian turbulent kinetic energy,  $TKE_E = \frac{1}{2} \overline{{\mathbf u^{\prime}}\cdot{\mathbf u^{\prime}}}$, a transfer term appears of the form $\partial_t TKE_E =  - \overline{{\mathbf u^{\prime}} w^{\prime}} \cdot \partial_z \overline{{\mathbf u^S}}$ \cite[]{McWilliams1997}. 
It is interpreted as a transfer from surface wave energy into $TKE_E$ and originates from the Stokes term in the modified pressure $p^*$. The transfer term was also derived from rapid distortion theory
\cite[]{Teixeira2002} and from generalised Lagrangian mean theory \cite[]{Ardhuin2006}. Apart from breaking waves, this transfer often dominates the $TKE$ budget in the ocean mixed layer \cite[]{belcher2012global} and is associated with 
Langmuir circulation, and thus may also play an important role in the ocean energy cycle.
It is the aim of this study to integrate 
all such energy transfers into a meaningful and consistent Lagrangian  framework.

In Section 2 we discuss the mean kinetic energy equations in this Lagrangian framework which allow for a closed energy budget with well known energy transfer terms. Large eddy simulations are used to visualise some important energy transfers in idealised experiments in Section 3.
As the model community starts to become aware of energy  consistency (e.g. \cite{Eden2014}), we  discuss in Section 4, how such a consistent framework could be realised in a general circulation ocean model coupled to a surface wave model.        

\section{Energy budgets}

\subsection{Mean kinetic energy}
 
The surface-wave averaged Boussinesq momentum Eq.~(\ref{eq1})  can be rewritten in a mathematical identical form (SFK16):
\begin{equation}\label{eq2}
\partial_t {\mathbf u} + ({\mathbf u^L} \cdot \nabla) {\mathbf u} +  {\mathbf f} \times  {\mathbf u^L}  = b {\mathbf z} - \nabla p  -  u^L_i \nabla  u^S_i  + {\mathbf D_u} 
\end{equation}
Here, ${\mathbf u}$ is the wave-averaged Eulerian velocity, ${\mathbf u^L} = {\mathbf u} + {\mathbf u^S} $  the Lagrangian velocity, i.e.  the sum of the Eulerian velocity  ${\mathbf u}$ plus the Stokes drift ${\mathbf u^S}$
with components $u^L_i$ and $u^S_i$.
Surface wave effects enter the momentum equations via the Stokes drift and modify the advection and  the Coriolis term, i.e. the Coriolis-Stokes force adds to the "traditional" Coriolis force.  
The third term on the r.h.s. is the Stokes shear force and is responsible for Langmuir turbulence.  Note, that  Einstein summation convention is used here. 
For now we consider only molecular dissipation acting on the Lagarangian motion, so that ${\mathbf D_u}  = \mu  {\mathbf \nabla^2} {\mathbf u^L}$, with $\mu$
 being the molecular viscosity.

A mean kinetic energy equation  for the Eulerian velocity,  $MKE_{E} = \frac{1}{2} \overline{u}_i \, \overline{u}_i$, can be derived by multiplying equation (\ref{eq2}) by $\cdot \overline{{\mathbf u}}$, followed by a suitable averaging denoted by an overbar. The averaging should satisfy $\overline{\nabla_{\alpha} A} = \nabla_{\alpha} \overline{A}$ for $\alpha=(t,x,y,z)$, $\overline{\overline{A}} = \overline{A}$, and
$\overline{\overline{A} B} = \overline{A}\; \overline{B}$ for
any quantities $A$ and $B$. Suitable methods are therefore ensemble and horizontal averages. Here we choose horizontal averaging, as we will later show model results in a horizontally periodic domain. The velocity $ {\mathbf u}$ can be therefore split into
$ {\mathbf u} =  \overline{{\mathbf u}} + {\mathbf u}^{\prime}$, where the turbulent velocity ${\mathbf u}^{\prime}$ is the deviation from the mean velocity $\overline{{\mathbf u}}$. 
In the same way we split
${\mathbf u^L} $ and ${\mathbf u^S}$. As all our velocities are phase-averaged with respect to surface gravity waves, the above horizontal averaging relies on the non-trivial assumption that turbulent quantities are de-correlated from the wave phase. This might be especially problematic with respect to  ${\mathbf u}^{S \prime}$, however, we here follow SFK16 and keep this assumption for now.

$\partial_t MKE_{E}$  is then given by:
\begin{equation}\label{eq3}
\begin{split}
\frac{\partial} {\partial t} {MKE_{E}} +  \underbrace{\overline{u^L_j} \frac {\partial} {\partial x_j} MKE_{E}}_{1}  +  \underbrace{\vphantom{\frac {\partial K_{ME}} {\partial x_j}}\frac {\partial} {\partial x_j} (\overline{u_j}  \, \overline{p}) }_{2} +  \underbrace{\frac {\partial} {\partial x_j} \left(\overline{u_i}  \, \overline{u_i^{\prime} u_j^{L \prime}} \right)}_{3}  -  \underbrace{ \frac {\partial} {\partial x_j} \left(\mu  \,  \overline{u_i} \frac{ \partial  \overline{u^L_i}}{\partial x_j}  \right)}_{4} = \\\  \underbrace{ \overline{u_i^{\prime} u_j^{L \prime}} \frac {\partial  \overline{u_i }} {\partial x_j}}_{5} + \underbrace{\vphantom{ \frac {\partial  \overline{u_i} } {\partial x_j}} \delta_{i,3} \overline{b}  \overline{u_i}}_{6}  - \underbrace{\vphantom{ \frac {\partial  \overline{u_i} } {\partial x_j}} \mu  \frac {\partial  \overline{u_i} } {\partial x_j}  \frac {\partial  \overline{u^L_i} } {\partial x_j}}_{7} - \underbrace{\vphantom{ \frac {\partial  \overline{u}_i } {\partial x_j}} \epsilon_{ijk} \overline{u_i} f_j \overline{u^S_k}}_{8} - \underbrace{\vphantom{ \frac {\partial  \overline{u}_i } {\partial x_j}}\overline{u}_i \overline{u^L_j}   \frac {\partial  \overline{u^S_j}} {\partial x_i}}_{9} -  \underbrace{\vphantom{ \frac {\partial  \overline{u}_i } {\partial x_j}} \overline{u}_i \overline{u^{L \prime}_j   \frac {\partial  u^{S \prime}_j } {\partial x_i}}}_{10}
\end{split} 
\end{equation}
Eq.~(\ref{eq3}) distinguishes transport terms on the l.h.s. from  exchange terms on the r.h.s.
The terms 1 to 4 are advection of $MKE_{E}$, work done by pressure, transport by the Reynolds stresses, and transport by viscous stresses, respectively. 
These terms redistribute mean kinetic energy. 
Note, that here $MKE_{E}$ is advected by the Lagrangian mean velocity $\overline{{\mathbf u^L}}$ and that the Reynolds stresses combine the deviations from the Eulerian and Lagrangian mean, i.e. $ \overline{u_i^{\prime} u_j^{L \prime}}$. 
Term 5 gives the exchange with  turbulent kinetic energy,  followed by the exchange with mean potential energy (term 6). 
The molecular dissipation of $MKE_{E}$ is given by term 7.  The dissipation is not positive definite, which depends whether the dissipation acts on the Eulerian or the Lagrangian velocity. The decision has consequences of either having a non-positive definite dissipation in the Eulerian energy or helocity budget \cite[]{Holm1996}. We assume that the molecular dissipation acts on the Lagrangian velocity, which leads to a positive definite dissipation in the Lagrangian energy budget introduced further below. 

The term 8 in Eq.~(\ref{eq3}) is the work done by the Coriolis-Stokes force. Finally, we have the somewhat unfamiliar terms 9 and 10 which both originate from the Stokes shear force.  Eq.~(\ref{eq3}) is first given in a form which is similar to SFK16, who already derived the terms 8 to 10. These terms  are  absent in an energy equation without the presence of surface waves and therefore represent energy exchanges with surface wave energy (SFK16). To our knowledge, the full $MKE_E$ equation in the current form is here presented for the first time.

Eq.~(\ref{eq3}) can be reformulated in the more familiar form:
\begin{equation}\label{eq4}
\begin{split}
\frac{\partial} {\partial t} {MKE_{E}} +  \underbrace{\overline{u_j} \frac {\partial} {\partial x_j} MKE_{L}}_{T1^E}  +  \underbrace{\vphantom{\frac {\partial K_{ME}} {\partial x_j}}\frac {\partial} {\partial x_j} \left(\overline{u_j}  \, \overline{p}\right) }_{T2^E} +  \underbrace{\frac {\partial} {\partial x_j} \left(\overline{u_i}  \, \overline{u_i^{L \prime} u_j^{L \prime}} \right)}_{T3^E} -  \underbrace{ \frac {\partial} {\partial x_j} \left(\mu  \,  \overline{u_i} \frac{ \partial  \overline{u^L_i}}{\partial x_j}  \right)}_{T4^E} = \\\  \underbrace{ \overline{u_i^{L \prime} u_j^{L \prime}} \frac {\partial  \overline{u_i }} {\partial x_j}}_{E1^E} + \underbrace{\vphantom{ \frac {\partial  \overline{u}_i } {\partial x_j}} \delta_{i,3} \overline{b}  \overline{u_i}}_{E2^E}  -\underbrace{\vphantom{ \frac {\partial  \overline{u_i} } {\partial x_j}} \mu  \frac {\partial  \overline{u_i} } {\partial x_j}  \frac {\partial  \overline{u^L_i} } {\partial x_j}}_{E3^E} - \underbrace{\vphantom{ \frac {\partial  \overline{u}_i } {\partial x_j}} \epsilon_{ijk} \overline{u_i} f_j \overline{u^S_k}}_{E^4}
\end{split} 
\end{equation}

The reformulation allows for an easier interpretation of the individual terms and is given here for the first time. The labelling of the terms in Eq.~(\ref{eq4}) distinguishes transport terms $T$ from exchange terms $E$.  The advection term $T1^E$ in  Eq.~(\ref{eq4}) is now expressed as the advection of  Lagrangian mean kinetic energy,  $MKE_L = \frac{1}{2} \overline{{\mathbf u^L}} \cdot \overline{{\mathbf u^L}}$, advected by the eulerian velocity $\overline{{\mathbf u}}$. The Reynolds stresses in term $T3^E$ and $E1^E$ are now the deviations from Lagrangian mean velocities. The unfamiliar terms 9 and 10 of  Eq.~(\ref{eq3}) are absorbed into the more familiar terms $T1^E$, $T3^E$ and $E1^E$ of Eq. (\ref{eq4}), i.e. in the latter form they can be interpreted as either transport terms or in case of $E1^E$ as an exchange with turbulent kinetic energy.

Adding $ \partial_t {\mathbf u^S} $ on both sides of Eq.~(\ref{eq2})  allows to write a tendency equation for the Lagrangian velocity \cite[]{Holm1996}:
\begin{equation}\label{eq5}
\partial_t {\mathbf u^L} + ({\mathbf u^L} \cdot \nabla) {\mathbf u^L} +  {\mathbf f} \times  {\mathbf u^L}  = b {\mathbf z} - \nabla p  - {\mathbf u^L} \times (\nabla \times {\mathbf u^S}  )   + {\mathbf D_u} + \partial_t {\mathbf u^S} 
\end{equation}

The momentum equation Eq. (\ref{eq5}) will be the cornerstone, on which our Lagrangian energy budget and suggested model framework relies. In contrast to Eq. (\ref{eq2}), velocities are fully written in terms of ${\mathbf u^L}$ and ${\mathbf u^S}$, i.e. the Eulerian velocity is absent. The term $- {\mathbf u^L} \times (\nabla \times {\mathbf u^S}) =  - u^L_i \nabla  u^S_i + ({\mathbf u^L} \cdot \nabla) {\mathbf u^S}$ contains the Stokes shear term and the advection of ${\mathbf u^S}$.
A temporal change in the Stokes drift $\partial_t {\mathbf u^S}$ can be interpreted as a forcing term for the Lagrangian velocity. 

Multiplying Eq.~(\ref{eq5}) by $\cdot \overline{ \mathbf u^L}$ and averaging leads then to a tendency equation for the Lagrangian mean kinetic energy $MKE_{L} = \frac{1}{2} \overline{u^L_i} \, \overline{u^L_i} $:

\begin{equation}\label{eq6}
\begin{split}
\frac{\partial} {\partial t} {MKE_{L}} +  \underbrace{\overline{u_j^L} \frac {\partial} {\partial x_j} MKE_{L}}_{T1^L}  +  \underbrace{\vphantom{\frac {\partial K_{ME}} {\partial x_j}}\frac {\partial} {\partial x_j} \left(\overline{u_j^L}  \, \overline{p} \right) }_{T2^L} +  \underbrace{\frac {\partial} {\partial x_j} \left(\overline{u_i^L}  \, \overline{u_i^{L \prime} u_j^{L \prime}} \right)}_{T3^L}  -  \underbrace{ \frac {\partial} {\partial x_j} \left(\nu \frac{ \partial }{\partial x_j} {MKE_{L}}  \right)}_{T4^L}  = \\\   \underbrace{ \overline{u_i^{L \prime} u_j^{L \prime}} \frac {\partial  \overline{u_i^L }} {\partial x_j}}_{E1^L} + \underbrace{\vphantom{ \frac {\partial  \overline{u}_i } {\partial x_j}} \delta_{i,3} \overline{b}  \overline{u_i^L}}_{E2^L}  - \underbrace{\vphantom{ \frac {\partial  \overline{u}_i } {\partial x_j}} \mu  \frac {\partial  \overline{u^L_i} } {\partial x_j}  \frac {\partial  \overline{u^L_i} } {\partial x_j}}_{E3^L} + \underbrace{ \vphantom{\frac {\partial  \overline{u}_i } {\partial x_j}} \overline{u^L_i} \frac{ \partial \overline{u^S_i}}{\partial t}}_{E7^L} 
\end{split} 
\end{equation}

The labelling of Eq.~(\ref{eq6}) follows Eq.~(\ref{eq4}), i.e. the number indicates the  physical interpretation of the term. A short version of Eq.~(\ref{eq6}), without the terms $T3^L$, $E1^L$ and $E2^L$, is also given in \cite{Holm1996}.
We first notice that apart from the new forcing term $E7^L$, Eq.~(\ref{eq6}) represents the standard  textbook form of a mean kinetic energy equation (e.g. \cite{Olbers2012}, their Eq. 11.63), but here purely expressed using the Lagrangian velocity ${\mathbf u^L}$. The Coriolis force is absent in the energy budget for $MKE_{L}$. We further notice that the exchange term  with turbulent Lagrangian energy, $E1^L$, includes $\overline{u_i^{\prime} u_j^{\prime}} \partial_{x_j}  \overline{u_i^S}$, which is missing in $E1^E$ of Eq.~(\ref{eq4}). 

We propose energy equation Eq.~(\ref{eq6}) for $MKE_L$ to be used for interpretation and quantification of energy transfers and budget in the surface wave effected ocean. Eq.~(\ref{eq6}) will be exploited in section 4, in order to establish an energetically consistent coupling between a surface wave and ocean model.

$MKE_L$ can be split into different energy compartments:
\begin{equation}\label{eq7}
MKE_L = \frac{1}{2} ( \overline{{\mathbf u}}  +  \overline{{\mathbf u^S}})  \cdot ( \overline{{\mathbf u}} +  \overline{{\mathbf u^S}})  =  MKE_E + MKE_S +  MKE_{ES}
\end{equation}
$MKE_S =  \frac{1}{2}  \overline{{\mathbf u^S}}\cdot \overline{{\mathbf u^S}}$ is the kinetic energy in the Stokes drift, its evolution is solely given by our prescribed forcing, i.e.  $\partial_t  MKE_S = \overline{{\mathbf u^S}}\cdot \partial_t \overline{{\mathbf u^S}}$.  $MKE_{ES} =   \overline{{\mathbf u}}\cdot \overline{{\mathbf u^S}}$  is some mixed Eulerian velocity/ Stokes drift energy. By construction $MKE_{ES}$ is not always positive. A possible negative energy already shows that the split of $MKE_L$ into the different compartments is a purely mathematical construct which 
cannot be based on physical arguments. However,
we follow this route for a moment here.
A tendency equation for $MKE_{ES}$ can be derived by multiplying   Eq.~(\ref{eq2})  by 
 $\cdot {\mathbf u^S}$ followed by averaging, and is given by
\begin{equation}\label{eq8}
\begin{split}
\frac{\partial} {\partial t} {MKE_{ES}} +  \underbrace{\overline{u_j^S} \frac {\partial} {\partial x_j} MKE_{L}}_{T1^{ES}}  +  \underbrace{\vphantom{\frac {\partial K_{ME}} {\partial x_j}}\frac {\partial} {\partial x_j} \left(\overline{u_j^S}  \, \overline{p}\right) }_{T2^{ES}} +  \underbrace{\frac {\partial} {\partial x_j} \left(\overline{u_i^S}  \, \overline{u_i^{L \prime} u_j^{L \prime}} \right)}_{T3^{ES}}    -  \underbrace{ \frac {\partial} {\partial x_j} \left(\mu  \,\overline{u^S_i} \frac{ \partial  \overline{u^L_i} }{\partial x_j}   \right)}_{T4^{ES}} = \\\   \underbrace{ \overline{u_i^{L \prime} u_j^{L \prime}} \frac {\partial  \overline{u_i^S }} {\partial x_j}}_{E1^{ES}} + \underbrace{\vphantom{ \frac {\partial  \overline{u}_i } {\partial x_j}} \delta_{i,3} \overline{b}  \overline{u_i^S}}_{E2^{ES}}  -\underbrace{\vphantom{ \frac {\partial  \overline{u}_i } {\partial x_j}} \mu  \frac {\partial  \overline{u^S_i} } {\partial x_j}  \frac {\partial  \overline{u^L_i} } {\partial x_j}    }_{E3^{ES}} - \underbrace{\vphantom{ \frac {\partial  \overline{u}_i } {\partial x_j}} \epsilon_{ijk} \overline{u^S_i} f_j \overline{u_k}}_{E4^{ES}} + \underbrace{ \vphantom{\frac {\partial  \overline{u}_i } {\partial x_j}} \overline{u_i} \frac{ \partial \overline{u^S_i}}{\partial t} }_{E7^{ES}} 
\end{split} 
\end{equation}

By construction, the individual terms of Eq.~(\ref{eq8}) can be added to the terms of the  tendency equations for $MKE_{E}$  and $MKE_{S}$ to give the tendency equation for $MKE_{L}$.  The triple product of term $E4^{ES}$ in Eq.~(\ref{eq8}) corresponds to the triple product of term $E4^{E}$ in Eq.~(\ref{eq4}) only with opposite sign, i.e. the work done by the Coriolis-Stokes force exchange energy between $MKE_{ES}$ and $MKE_{E}$. As both are part of the Lagrangian energy  $MKE_{L}$, the work done by the Coriolis-Stokes force does not appear in the budget for $MKE_{L}$.

A physical interpretation and/or justification for $MKE_{ES}$ is lacking, and so is any exchange between $MKE_{ES}$ and $MKE_{E}$. We therefore encourage to use the the Lagrangian energy  $MKE_{L}$ budget, were such an energy transfer term based on the  fictitious Coriolis force is absent. 

\subsection{Turbulent kinetic energy}

An equation for the turbulent Lagrangian kinetic energy,  $TKE_{L} = \frac{1}{2} \overline{{\mathbf u^{L \prime }} \cdot {\mathbf u^{L \prime}}}$ can be derived by multiplying  Eq.~(\ref{eq5}) by $\cdot  \mathbf u^{L \prime}$ and averaging:
\begin{equation}\label{eq9}
\begin{split}
\frac{\partial} {\partial t} {TKE_{L}} +  \underbrace{\overline{u_j^L} \frac {\partial} {\partial x_j} TKE_{L}}_{T1^{L \prime}}  +  \underbrace{\vphantom{\frac {\partial K_{ME}} {\partial x_j}}\frac {\partial} {\partial x_j} \overline{u_j^{L \prime}  \, p^{\prime}} }_{T2^{L \prime}} +  \underbrace{\frac {\partial} {\partial x_j} \left( \frac{1}{2} \overline{u_j^{L \prime}  \, u_i^{L \prime} u_i^{L \prime}} \right)}_{T3^{L \prime}}  -  \underbrace{ \frac {\partial} {\partial x_j} \left(\mu \frac{ \partial }{\partial x_j} {TKE_{L}}  \right)}_{T4^{L \prime}} = \\\   - \underbrace{ \overline{u_i^{L \prime} u_j^{L \prime}} \frac {\partial  \overline{u_i^L }} {\partial x_j}}_{E1^{L \prime}} + \underbrace{\vphantom{ \frac {\partial  \overline{u}_i } {\partial x_j}} \delta_{i,3} \overline{b^{\prime} u_i^{\prime L}}}_{E2^{L \prime}}  -\underbrace{\vphantom{ \frac {\partial  \overline{u}_i } {\partial x_j}} \mu \overline{ \frac {\partial  u^{L \prime}_i}  {\partial x_j}  \frac {\partial  u^{L \prime}_i}  {\partial x_j}}}_{E3^{L \prime}} + \underbrace{ \vphantom{\frac {\partial  \overline{u}_i } {\partial x_j}}  \overline{u^{L \prime}_i \frac{ \partial u^{S \prime}_i}{\partial t} }}_{E7^{L \prime}} 
\end{split} 
\end{equation}

 The tendency equation for $TKE_{L}$ represents again the textbook form of a turbulent kinetic energy equation (e.g. \cite{Olbers2012}, their Eq. 11.62), with the addition of a forcing term $E7^{L \prime}$. The transport terms $T1^{L \prime}$ to $T4^{L \prime}$ are Lagrangian advection of $TKE_{L}$, work done by pressure fluctuation, transport by Reynolds-, and by viscous stresses, respectively. Term $E1^{L \prime}$ represents the exchange term with $MKE_{L}$ and corresponds to term $E1^L$ of Eq.~(\ref{eq6}). Term $E2^{L \prime}$ gives the exchange with turbulent potential energy, and $E3^{L \prime}$  describes the molecular dissipation of $TKE_{L}$.

 Similar to the mean kinetic energy, we can split $TKE_{L}$ into different compartments , i.e.  $TKE_{L} = TKE_{E}  + TKE_{S} + TKE_{ES}$. $TKE_S =  \frac{1}{2}  \overline{{\mathbf u^{S \prime}}\cdot {\mathbf u^{S \prime}}}$ is the kinetic energy in the Stokes drift fluctuations, its evolution is given by the forcing term, i.e.  $\partial_t  TKE_S = \overline{{\mathbf u^{S \prime}}\cdot \partial_t {\mathbf u^{S \prime}}}$.  $TKE_{ES} = \overline{{\mathbf u^{ \prime}}\cdot {\mathbf u^{S \prime}}}$ is a unfamiliar mixed turbulent Eulerian/Stokes energy which is not positive definite.
 Finally, $TKE_{E} = \frac{1}{2}  \overline{{\mathbf u^{ \prime}}\cdot {\mathbf u^{\prime}}}$  is the Eulerian turbulent kinetic energy, its evolution is given by:
 
\begin{equation}\label{eq10}
\begin{split}
\frac{\partial} {\partial t} {TKE_{E}} +  \underbrace{ \vphantom{ \left(\nu  \,  \overline{u_i^{\prime} \frac{ \partial u^{L \prime}_i}{\partial x_j}}  \right)}\overline{u_j} \frac {\partial} {\partial x_j} TKE_{L}}_{T1^{E \prime}}  +  \underbrace{\vphantom{ \left(\nu  \,  \overline{u_i^{\prime} \frac{ \partial u^{L \prime}_i}{\partial x_j}}  \right)} \frac {\partial} {\partial x_j} \overline{u_j^{\prime}  \, p^{\prime}} }_{T2^{E \prime}} +  \underbrace{\vphantom{ \left(\nu  \,  \overline{u_i^{\prime} \frac{ \partial u^{L \prime}_i}{\partial x_j}}  \right)}\frac {\partial} {\partial x_j} \left( \frac{1}{2} \overline{u_j^{\prime}  \, u_i^{L \prime} u_i^{L \prime}} \right)}_{T3^{E \prime}}   -  \underbrace{ \frac {\partial} {\partial x_j} \left(\mu  \,  \overline{u_i^{\prime} \frac{ \partial u^{L \prime}_i}{\partial x_j}}  \right)}_{T4^{E \prime}}  = \\\ 
  -   \underbrace{ \overline{u_i^{ \prime} u_j^{L \prime}} \frac {\partial  \overline{u_i^L}} {\partial x_j}}_{E1^{E \prime}} + \underbrace{\vphantom{ \frac {\partial  \overline{u}_i } {\partial x_j}} \delta_{i,3} \overline{b^{\prime} u_i^{\prime }}}_{E2^{E \prime}}  - \underbrace{\vphantom{ \frac {\partial  \overline{u}_i } {\partial x_j}}  \mu \overline{ \frac {\partial  u^{ \prime}_i}  {\partial x_j}  \frac {\partial  u^{L \prime}_i}  {\partial x_j}}}_{E3^{E \prime}} - \underbrace{\vphantom{ \frac {\partial  \overline{u}_i } {\partial x_j}} \epsilon_{ijk} \overline{u_i^{\prime } f_j u^{S \prime}_k}}_{E4^{E \prime}} 
\end{split} 
\end{equation}
The term $E4^{E \prime}$ represents work done by the fluctuating part of the Coriolis-Stokes force. 
It does not show up in the $TKE_{L}$ budget, as it exchanges energy with $TKE_{ES}$ in the same way as its mean part exchanges energy between $MKE_{E}$ and $MKE_{ES}$. The shear production term $E1^{E \prime}$ includes the full Lagrangian shear, i.e. it represents only partly an exchange  with  $MKE_{E}$ given by the Eulerian shear production, but additionally exchange energy with $MKE_{ES}$ given by the Stokes shear production term $E1^{ES}$.

\subsection{Lagrangian vs. Eulerian energy budget}

\begin{figure}[hh]
  \centerline{\includegraphics[width=40pc,angle=0]{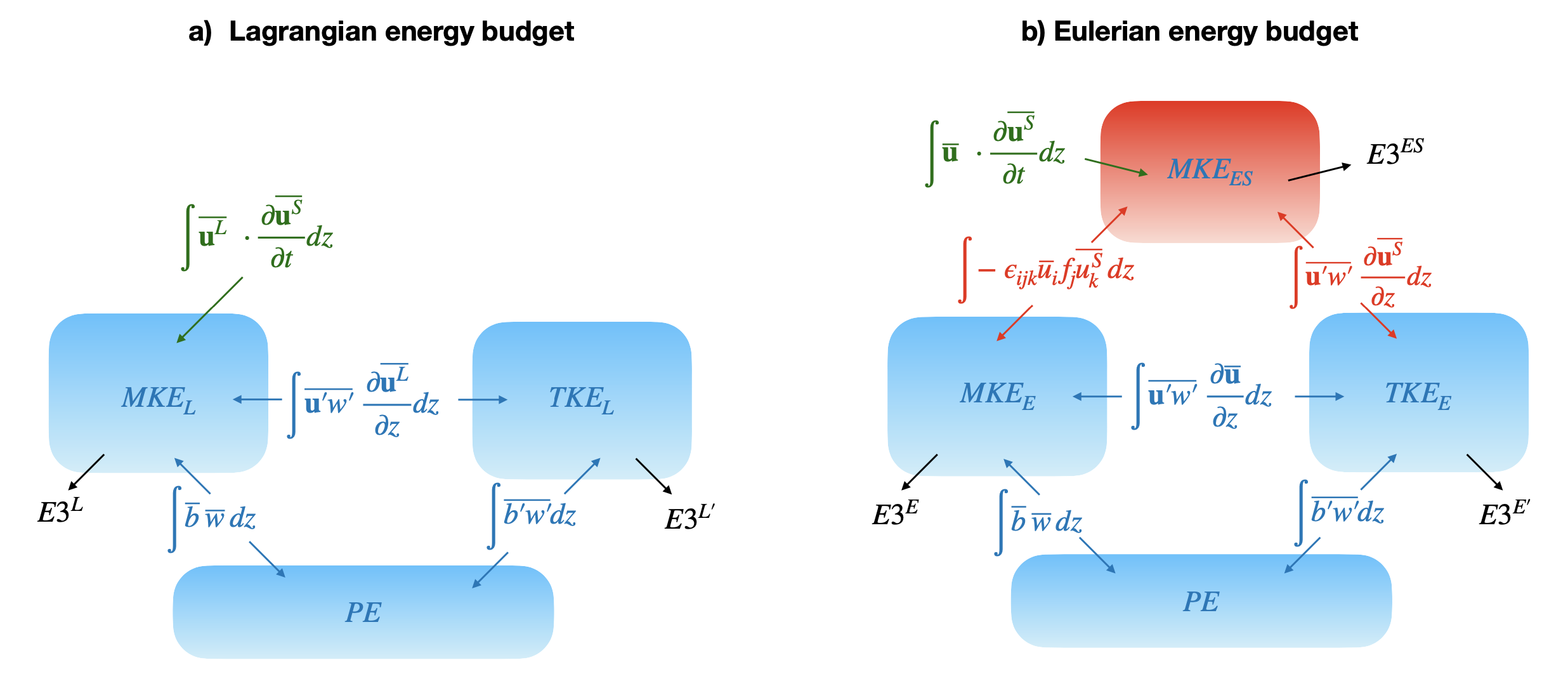}}
   \caption{Schematic of the energy exchanges between different compartments (boxes) for a) the  Lagrangian and b) the Eulerian framework. Blue transfer terms give transfers within the mechanical energy budget of the ocean. Green transfer terms give exchanges with surface wave energy through temporal changes in the Stokes drift, and black transfer terms give exchanges with internal energy through molecular dissipation (E3 terms). Finally, red transfer terms
   give exchanges with the physically difficult to interpret compartment $MKE_{ES}$.
   Energy fluxes at the vertical boundaries 
   and lateral flux divergences are not shown.   
   } 
   \label{fig1}
\end{figure}

In order to simplify the comparison between Lagrangian and Eulerian
framework, we consider the plausible assumption that ${\mathbf u^{S \prime}} = 0$. The result is that $TKE_{L}$ and  $TKE_{E}$, and their corresponding tendency Eq.~(\ref{eq9}) and (\ref{eq10}) are identical. 
Schematics of the vertically integrated energy budgets which then follow are given in Fig.~(\ref{fig1}). 
Our advocated Lagrangian framework has a closed mechanical energy budget. Compartments and the well known energy transfer terms  are given in black in Fig.~(\ref{fig1}a). Exchange terms with other energy compartments are colored in blue. Exchanges with internal energy through molecular dissipation are given by the terms $E3^L$ and  $E3^{L ^\prime}$ of Eq.~(\ref{eq6}) and 
Eq.~(\ref{eq9}), respectively.  The only term, which exchange energy with  "external" surface wave energy in this framework, is given by the forcing term $\int \overline{\mathbf{u^L}} \cdot \partial_t \overline{\mathbf{u^S}}\;dz$. Note, that the term "external" surface wave energy is used here, as the inclusion of the Stokes drift in $MKE_L$ represents already some form of phase-averaged kinetic wave energy. Typically the wind stress, Stokes drift and near surface Lagrangian velocity are roughly aligned in the same direction, so that one can expect an overall energy transfer from the waves to oceanic motions by the forcing term, but exceptions seem possible.

Previous studies have interpreted the Eulerian mechanical energy budget (Fig.~(\ref{fig1}b)). The red energy transfer terms were interpreted as an energy exchange with surface wave energy (e.g. SFK16). However, surface wave energy equations showing these terms are lacking. A notable exception is the Stokes shear production term in the wave energy equation of \cite{Teixeira2002}. By splitting up  $MKE_L$ in the different compartments $MKE_E$, $MKE_{ES}$, and $MKE_S$, we showed that the red transfer terms can be interpreted as an exchange of $MKE_E$ and  $TKE_E$ 
with $MKE_{ES}$. The $MKE_{ES} =   \overline{{\mathbf u}}\cdot \overline{{\mathbf u^S}}$ compartment is given in red in (Fig.~(\ref{fig1}b)) because of its dubious physical meaning, for example, it is not positive definite.
As the definition of $MKE_{ES}$ includes the Eulerian velocity, there seems no reason to consider $MKE_{ES}$ as part of the surface wave energy. Any
physical interpretation of $MKE_{ES}$ or the associated energy exchanges
is very difficult. We therefore recommend to use the Lagrangian energy budget for quantification and interpretation of the energy transfers.
Using idealized numerical experiments, we will show in the next section, that a Lagrangian framework can lead to complete different results
in the energy transfers and budgets compared to the Eulerian framework.

In contrast to the energy transfer associated with the Coriolis-Stokes force, we do not wish to challenge the well-established interpretation of the Stokes shear production term $\overline{u_i^{\prime} u_j^{\prime}} \partial_{x_j}  \overline{u_i^S}$ as an exchange between  turbulence and surface wave energy. In the Lagrangian energy framework
this can be interpreted as follows: the Stokes shear production term
removes surface wave energy, which causes a change in the Stokes drift. 
The same amount of energy transfer is thus also contained in  ${\mathbf u^L} \cdot  \partial_t {\mathbf u^S}$. Note, however, that
changes in the surface wave energy due to wave growth and breaking are considered to be much larger than due to the Stokes shear production term \cite[]{Ardhuin2010}.

The  energy transfers through wave breaking and surface wave stresses can be interpreted as vertical boundary conditions for the transport terms labelled T3
and are not contained in the schematic in 
Fig.~(\ref{fig1}).
Implications for a large-scale numerical model framework  will be discussed in Section 4. 

\section{Numerical model experiments}

Although the results in this study are essentially analytically, we visualise some of our findings using a numerical model.
The model is designed to do large eddy simulations and is fully three
dimensional with lateral cyclic boundary conditions. At the vertical boundaries, we use no-flux boundary conditions. A sponge layer damps velocities near the bottom, in order to prevent reflection of internal waves.
The model  integrates the surface-wave averaged Boussinesq momentum given by Eq.~(\ref{eq2})  and a buoyancy equation  of the form: 
\begin{equation}\label{eq11}
\partial_t b + ({\mathbf u^L} \cdot \nabla) b = {D_b}
\end{equation}
The sub-grid scale closure follows \cite{Ducros1996}, with a turbulent eddy viscosity $\mu_t$ operating on the Eulerian velocity, so that  ${\mathbf D_u}  = {\mathbf \nabla}  \mu_t {\mathbf \nabla} {\mathbf (u^L-u^S})$. ${ D_b}  = {\mathbf \nabla}  \kappa_t {\mathbf \nabla} {b}$, with a turbulent eddy diffusivity  is $\kappa_t = \mu_t / Pr$, and  a Prandtl number of $Pr=0.7$. 

The wave forcing is given by a prescribed Stokes drift which corresponds  to  monochromatic,  uni-directional deep water wave:
 \begin{equation}\label{eq12}
 u^S_{eq} =  u_0^S \; \exp \left( \frac{z}{D_s} \right)
\end{equation}
$D_s=(2 k)^{-1}$ denotes the depth penetration scale, $k$ the wave number, and $u_0^{S}= a^2 \omega \,  k$ the surface Stokes drift,  $a$ as the amplitude, $g$ the gravitational acceleration, and $\omega=\sqrt{gk}$  the frequency. We choose typical swell conditions with a wave length of $\lambda = 100m$, an amplitude of $a=1.5m$, and a surface Stokes drift in positive x-direction of $u_0^{S}= 0.11 \frac{m}{s}$, leading to a depth penetration scale of $\sim 8m$.
For the growth of swell, we follow the analytical solution of  \cite{Wagner2021}:
 \begin{equation}\label{eq13}
 u^S(z,t) =  u^S_{eq}(z) \left[1 - \exp \left(\frac{-t^2}{2 T^2_w}\right)\right]
\end{equation}
Here, $ u^S_{eq}(z)$ is the equilibrated Stokes drift as given by Eq. (\ref{eq12}), and $T_w$ is a growth time scale, we chose  $T_w = 2 h$. 
The model uses a rigid-lid and we do not consider a wavy surface. A discussion of the Eulerian Stokes drift, which lies between the crests and troughs, is given by
 \cite{Brostrom2014}.

 \subsection{Laminar flow} 

The focus of the first experiment is on mean kinetic energies. The model is  initialised with a constant stratification of $N^2 = 5.0 \times 10^{-4} s^{-2}$. 
Although the Stokes shear term is included, the setup is chosen in a way, that the model does not generate Langmuir turbulence,  i.e. shear and Langmuir instabilities are not able to overcome the  initial stratification.  This also keeps the turbulent viscosity and diffusivity at the lower limit, which corresponds to molecular friction and diffusion. The results are basically one dimensional in the vertical, and all velocities can be considered as mean quantities. The lateral cyclic boundary conditions leads to a vanishing mean vertical velocity. The Coriolis frequency of ${\mathbf f} = f{\mathbf z} = 7.29  \times 10^{-5} s^{-1} $  corresponds to an inertial period of one day. 

In order to see how surface waves drive inertial oscillations theoretically, we make the following assumptions. We consider a linear, inviscid ocean away from lateral boundaries, so that no horizontal pressure gradient can be established on the considered scales. If we choose a horizontal Stokes drift of the form $u^S(z,t)$ the horizontal components of Eq.~(\ref{eq5}) simplify to
\begin{eqnarray}\label{eq14}
\partial_t {u^L} - fv &=& \partial_t {u^S}
\\
\label{eq15}
\partial_t {v^L} + fu +fu^S  &=& 0
\end{eqnarray}
A time dependent Stokes drift will lead to inertial oscillations as shown by  \cite{hasselmann1970wave}.  In steady state, the mass transport due to surface waves is exactly zero as the Stokes drift is balanced by the Eulerian anti-Stokes flow, i.e. $u^L,v^L=0$ and $u^S = -u$.  

\begin{figure}[h]
  \centerline{\includegraphics[width=30pc,angle=0]{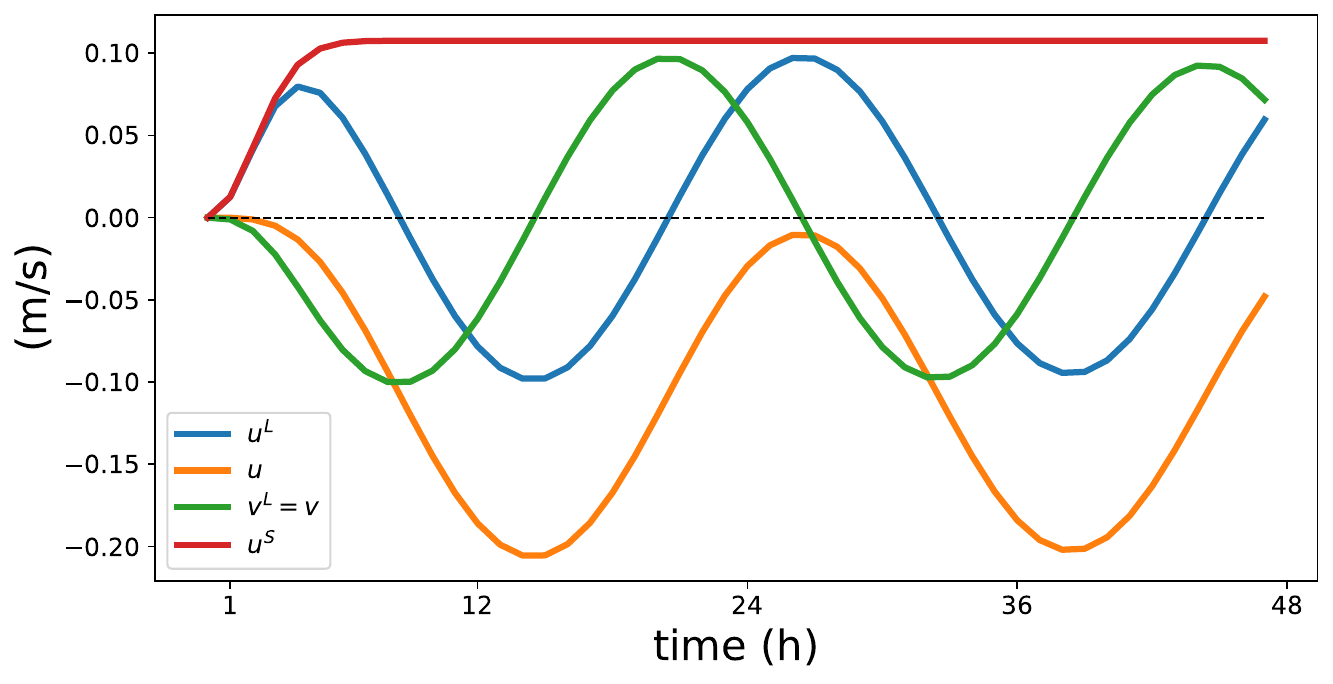}}
  \caption{Time evolution of surface velocities: Eulerian zonal velocity (orange), Lagrangian zonal velocity (blue),  Eulerian/Lagrangian meridional velocity (green), and zonal Stokes drift (red)} \label{fig2}
\end{figure}

Although the non-linear and inviscid  assumptions are not made in our model experiment, the results corresponds to Eqs.~(\ref{eq14}) and  (\ref{eq15}). Figure (\ref{fig2}) shows the different velocity compartments at the sea surface. The model is forced by the arrival of swell, i.e.  $\partial_t u^S$  in the first few hours according to Eq.~(\ref{eq13}) with a maximum forcing at $T_W=2h$. The forcing leads to nearly ideal inertial oscillations with vanishing time mean in the Lagrangian velocities. As the Stokes drift is only in zonal direction, the meridional component of the Eulerian and Lagrangian velocities are identical.  The amplitude of the inertial oscillations reaches close to the magnitude of the Stokes drift and
depends on the ratio between $T_w$ and the inertial period. If $T_w$ goes to zero, the amplitudes of the Stokes drift  and inertial oscillations will be equal. The Eulerian velocity is therefore always negative for a positive Stokes drift. If we average over one inertial period, the  Eulerian velocity will oppose the Stokes drift  $u = - u^S$,  known as  the anti-Stokes flow. The model therefore reproduces the findings of \cite{hasselmann1970wave}, and averaged over an inertial oscillation, the no net mass flux of \cite{Ursell1950}. 

In our simplified setting the Eulerian and Lagrangian
mean kinetic energy Eqs.~(\ref{eq4})  and  (\ref{eq6}) reduce to:
\begin{eqnarray}\label{eq16}
\partial_t {MKE_E} &=& - \overline{{\mathbf u}} \cdot f{\mathbf z} \times   \overline{{\mathbf u^S}}  
\\
\label{eq17}
\partial_t {MKE_L}  &=&   \overline{{\mathbf u^L}} \cdot \partial_t  \overline{{\mathbf u^S} }
\end{eqnarray}
Changes in $MKE_E$ are induced by the work done by the Coriolis-Stokes force. As noticed by \cite{Polton2009}, the term  ${\mathbf u} \cdot f{\mathbf z} \times  {\mathbf u^S}$  is a scalar product between a phase-averaged velocity and a phase-averaged non-linear momentum term (wave Reynolds stress), where the latter only gives $f{\mathbf z} \times  {\mathbf u^S}$ after phase-averaging. Therefore, the energy equation should be derived before phase-averaging, as wave correlated terms could give rise to an additional term in the energy budget. We checked that for our monochromatic wave, to find out that these contribution can be safely neglected here. In the tendency equation for the $MKE_L$ 
no Coriolis-Stokes force appears and the evolution is dependent on the forcing, i.e. $\overline{{\mathbf u^L}} \cdot  \partial_t \overline{{\mathbf u}^S}$. As outlined in the previous section $MKE_L = MKE_E + MKE_{ES} +MKE_{S}  $.
The two remaining compartments are given by 
\begin{eqnarray}\label{eq18}
\partial_t MKE_{ES}  &=& \overline{{\mathbf u}} \cdot f{\mathbf z} \times   \overline{{\mathbf u^S}}   +\overline{{\mathbf u}} \frac{\partial   \overline{{\mathbf u^S}}}{\partial t}
\\
\label{eq19}
\partial_t MKE_{S}   &=& \overline{ {\mathbf u^S}} \frac{\partial  \overline{{\mathbf u^S}}}{\partial t}
\end{eqnarray}
 
\begin{figure}[hh]
  \centerline{\includegraphics[width=30pc,angle=0]{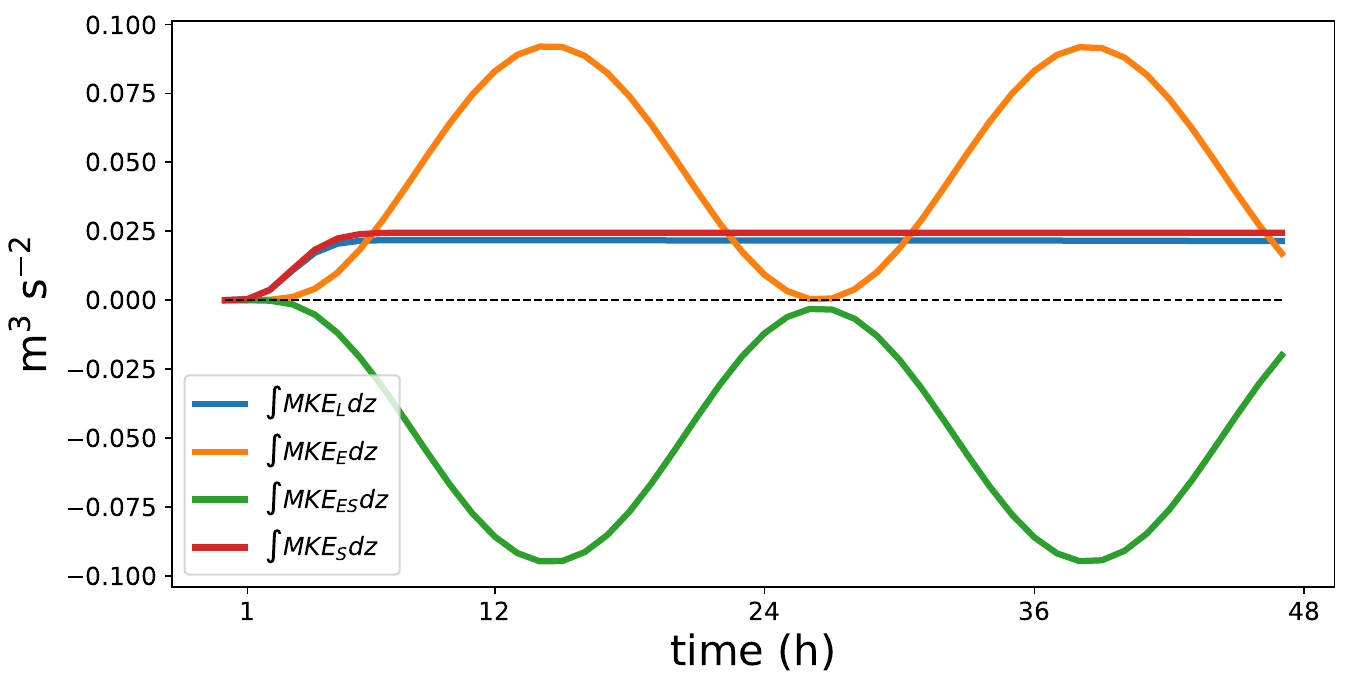}}
  \caption{Time evolution of vertically integrated mean kinetic energy compartments: Eulerian mean kinetic energy  $MKE_E$ (orange), Lagrangian mean kinetic energy  $MKE_L$ (blue),  the mixed  kinetic energy  $MKE_{ES}$ (green), and kinetic energy of the  Stokes drift $MKE_S$ (red). See text for details and definitions.} \label{fig3}
\end{figure}
 
 The vertically integrated energy budgets for the compartments  are shown in Fig.~(\ref{fig3}) . 
 The budget for $MKE_E$ shows strong undulations during an inertial cycle. Averaged over an inertial period the budget is  0.46 $m^3s^{-2}$ and much higher than for the Lagrangian energy $MKE_L$, which is 0.22 $m^3s^{-2}$.  
 The exchange between $MKE_{E}$ and $MKE_{ES}$ through the Coriolis-Stokes term is the dominant signal. The sum of the $MKE_E$ and the $MKE_{ES}$ budget is still slightly negative as it is determined by the negative  contribution of the r.h.s. of Eq.~(\ref{eq18}). 
 As no further energy exchange terms are given, the difference in the budget between  $MKE_{E}$ and $MKE_{ES}$ is identical to the difference between the two other compartments  $MKE_L$ and $MKE_S$.  $MKE_L$  gives the energy budget of the inertial oscillations (see also Fig.~(\ref{fig2})) in this experiment, which in a more complex setting will decrease over time mainly through form stress at the base of the mixed layer and dissipation (see \cite{Czeschel2019} and references therein). 
If all energy in the inertial oscillations is dissipated, $MKE_L$ would be zero and any movement of fluid particles would stop. However, $MKE_{E}$ would be still positive in such a steady state, as the Eulerian velocity would exactly oppose the Stokes drift. The physical interpretation of such an $MKE_{E}$ budget is difficult, as the energy cannot be transferred to, for example, $TKE$.

Multiplying the integrated mean energies by reference density, e.g.  $\rho_0 = 1000\, kg/ m^{3}$, allows a comparison with the total kinetic energy of the  surface waves of $E_{kin} = \frac{1}{4} g \rho_0 a^2 = 5518 \; J/m^2$, with $a=1.5 m$ being the amplitude of our prescribed swell. After the initial forcing period, $\int  \rho_0 \; MKE_{S} \;  dz \approx  \int  \rho_0 \;  MKE_{L} \;  dz  \approx 25 \; J/m^2$, i.e. it is just a small fraction of the total kinetic energy of the surface waves. $E_{kin}$  accounts for the full orbital motion, whereas $MKE_{S}$ accounts only for the Stokes drift, i.e. the deviation from a closed orbital loop. The $MKE_L$ budget of $\approx 25 \; J/m^2$ corresponds to the energy loss of $E_{kin}$ to inertial oscillation within the first $\approx 4$ hours as given by the r.h.s of Eq.~(\ref{eq17}). In contrast to the Eulerian energy budget, the Lagrangian energy budget therefore allows for a clear physical interpretation of the exchange terms.

\subsection{Turbulent flow} 

The impact of turbulence on surface waves and the associated Stokes drift is largely unknown. Applying the same eddy viscosity on the Stokes drift as used in mixing parameterizations for the upper ocean Eulerian currents seems not appropriate, as the energy loss would be much too strong \cite[]{Ardhuin2006}. 
A physical explanation for the different impact of turbulence
on the Stokes drift and the Eulerian current might be given by the overlapping time and spatial scales in wave dynamics and turbulence. For example, the time scale associated with Langmuir turbulence is often larger than the wave periods of typical wind waves, i.e. Langmuir turbulence might have no impact on such waves. The consequence for the energy budget would be that the exchange terms between $MKE_L$ and $TKE_L$ in Eqs.~(\ref{eq6}) and (\ref{eq9}), i.e. $\overline{u_i^{L \prime} u_j^{L \prime}} \frac {\partial  \overline{u_i^L }} {\partial x_j}$, do not share the same Reynolds stresses as they act on the Eulerian or the Stokes drift shear. The Lagrangian shear production term should be then reformulated as $\overline{u_i^{L \prime} u_j^{L \prime}}^E \frac {\partial  \overline{u_i }} {\partial x_j}+\overline{u_i^{L \prime} u_j^{L \prime}}^S \frac {\partial  \overline{u_i^S }} {\partial x_j}$ with $\overline{u_i^{L \prime} u_j^{L \prime}}^E \ne \overline{u_i^{L \prime} u_j^{L \prime}}^S$, here the different overbars denote different averaging scales. Parameterizing these Reynolds stresses would then demand different scale-dependent eddy viscosities.

However, the different Reynolds stresses are difficult to realise in models using phase-averaged equations, like the Craik-Leibovich equations. Such models typically prescribe the Stokes drift, and possible impacts of turbulence on the Stokes drift are neglected. This is achieved by an eddy viscosity that acts only on the Eulerian velocity and by neglecting the advection of Stokes drift.

We follow this approach here, and repeat the experiment from the laminar case but  with a uniform mixed layer of 50m ($N^2 =  0 \,s^{-2}$) on top of the stratified interior ($N^2 = 5.0 \times 10^{-4} s^{-2}$). We additionally  cool the ocean for 6 hours with 10 $Wm^{-2}$, in order to generate some initial turbulence. The surface wave forcing follows again Eq.~(\ref{eq13}) starting at t=0 after the cooling period, i.e. the model starts without mean kinetic energy, but has a vertically integrated turbulent kinetic energy of  $TKE_{L}  = 1.73 \times10^{-4} m^3s^{-2}$. $TKE_{L}$ equals $TKE_{E}$ in our set-up, as the Stokes drift is horizontally constant and ${\mathbf u^{S \prime}} = 0$. The large time and spatial scales of 
our swell forcing suggest that all generated turbulence act on the Stokes drift shear. As we apply lateral cyclic boundary conditions, all vertical mean velocities are zero, and the evolution of the mean horizontal velocities  are governed by
\begin{eqnarray}\label{eq20}
\partial_t \overline{{u^L}} - f\overline{v} &=& \partial_z \left(\overline{\mu_t \frac{\partial u}{\partial z}}-\overline{u^{\prime} w^{\prime}} \right) + \partial_t \overline{{u^S}}
\\
\label{eq21}
\partial_t \overline{{v^L}} + f\overline{u} +f\overline{u^S}  &=&  \partial_z \left(\overline{\mu_t \frac{\partial v}{\partial z}}-\overline{v^{\prime} w^{\prime}}\right)
\end{eqnarray}
here $\mu_t$ is the turbulent eddy viscosity acting on the Eulerian velocity. Note again, that $v^S=0$ in our experiment, which also gives $v^L=v$. In steady state, and assuming no-flux boundary conditions at the surface and somewhere in the stratified interior, only the vertical integrals of the Stokes drift and the anti-Stokes flow balances, so that $\int \overline{u^S} dz = - \int \overline{u} dz$ and $\int \overline{v} dz = 0$. This should be compared to the laminar case (Eq.~\ref{eq14} and \ref{eq15}), where the steady state solution was $\overline{u^S}={-u}$. Resolved and unresolved turbulence are therefore shaping the vertical profiles of the Eulerian velocities. The assumption that turbulence acts solely on the Eulerian velocities has therefore strong consequences on
the vertical gradients of $\overline{u^L}, \overline{u}$, and $\overline{u^S}$, and therefore also for the exchange between
$MKE_L$ and  $TKE_L$ as given by the Lagrangian shear production term.

\begin{figure}[hh]
  \centerline{\includegraphics[width=25pc,angle=0]{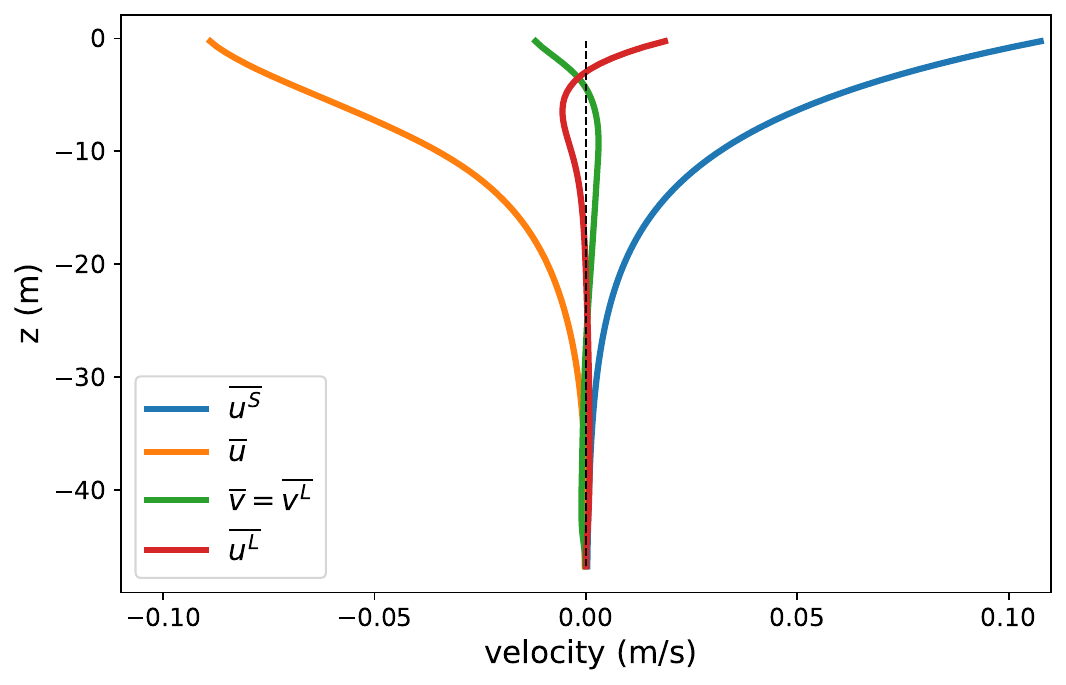}}
  \caption{ Vertical profiles of horizontal averaged velocities. The velocities are averaged over one inertial period, or from model hour 12 to 36.}
   \label{fig4}
\end{figure}

In our simulation, the vertical shear in the Stokes drift increases within the first $\sim 6$h due to the growing swell. The result is a  burst of turbulence driven by the developing Langmuir circulation. The $TKE$ budget is very similar to the findings of \cite{Wagner2021}  and is not repeated here. Although the turbulence is not in equilibrium in such a setup, the "quasi steady-state" velocity profiles from our turbulent experiment are given in Fig. \ref{fig4}. The shaping of the anti-Stokes flow $\overline{u}$ through turbulence is clearly visible. As predicted from Eq.~(\ref{eq20}) and Eq.~(\ref{eq21}) the vertical integrals
of $\overline{u}$ and $\overline{u^S}$ cancel each other, and $\overline{u^L}$ and $\overline{v}$ integrate to zero. The differences between the profiles of $\overline{u^S}$ and $(-) \,\overline{u}$ depend on the amount of turbulence, which is
rather weak in our experiment only driven by swell. 
For example, Langmuir turbulence driven by wind waves is usually stronger, and so are the differences between 
 $\overline{u}$ and $\overline{u^S}$. The differences become even larger, if other sources of turbulence also contribute.

\begin{figure}[hh]
  \centerline{\includegraphics[width=30pc,angle=0]{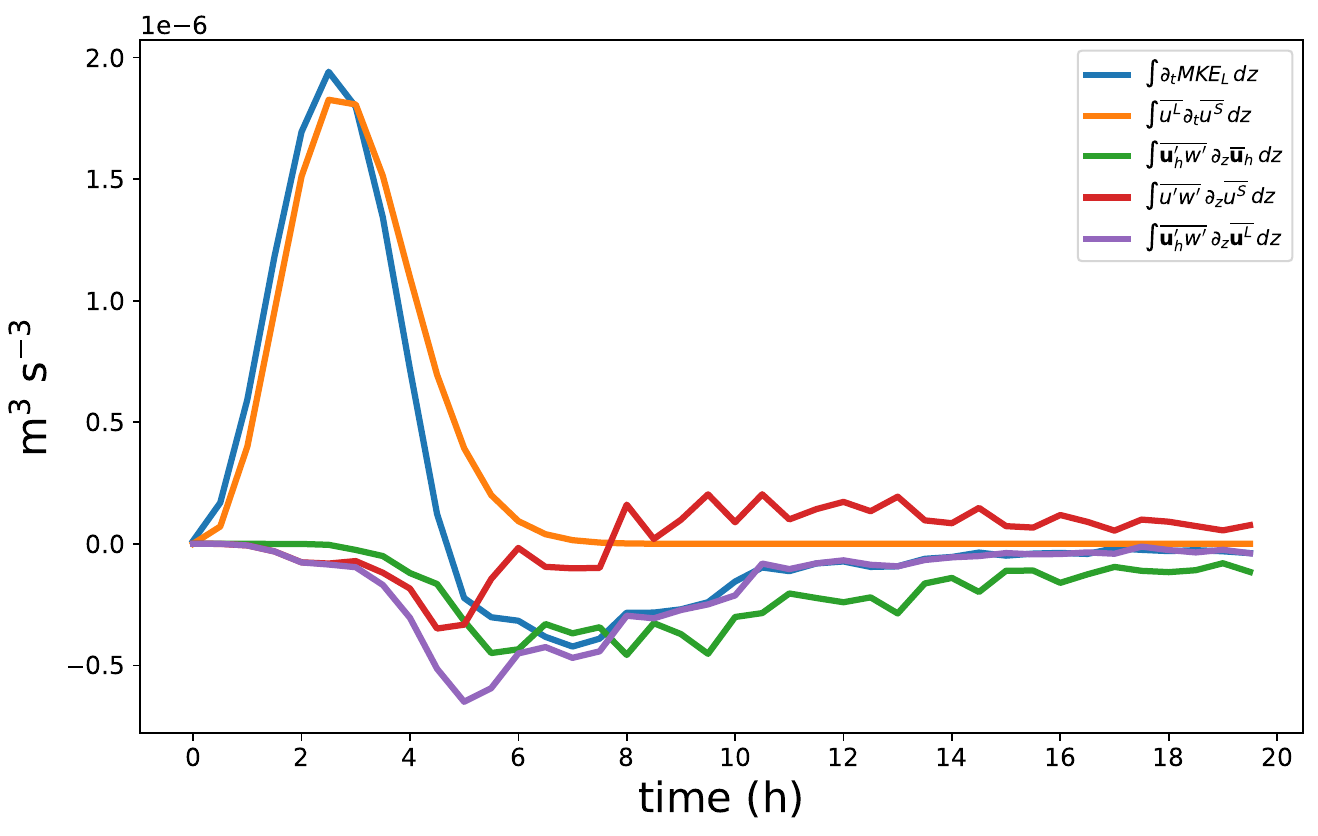}}
  \caption{Time evolution of vertically integrated mean kinetic energy tendency terms in $m^3/s^3$. Shown are temporal changes
  in $MKE_L$ (blue), the forcing due to a change in the Stokes drift (orange), the Lagrangian shear production term (purple) , which consists of the Eulerian  production term (shown in green) and the Stokes production term (red), and the  The vector $\mathbf{u}_h=(u,v)$ denotes the horizontal components.} 
    \label{fig5}
\end{figure} 

A detailed discussion of Langmuir turbulence driven by swell can be found in \cite{mcwilliams2014} and \cite{Wagner2021} and is not the scope of the present study. We here concentrate on the impact on $MKE_L$. The evolution in the vertically integrated tendency terms of  $MKE_L$ are given in Fig. \ref{fig5}. Similar to the laminar case, the temporal change in $MKE_L$ (blue) is initially  given by the forcing term due to temporal changes in the Stokes drift (orange). As in the laminar case, the term forces surface wave driven 
inertial oscillations. Around model hour three, Langmuir turbulence start to transfer energy from $MKE_L$ to $TKE_L$ as given by the Lagrangian shear production term (purple). After model hour 7 the change in $MKE_L$ is solely given by Lagrangian shear production term. The Lagrangian shear production consists of the Eulerian (green) and Stokes shear production (red). The Stokes shear production term changes sign after $\sim$ 9 hours. This is because the Reynolds stresses also changes sign, as they rotate with the Lagrangian mean flow,  which is effected by the inertial oscillation (see \cite{mcwilliams2014} for details). 
The  Eulerian  and Stokes shear production show some high frequent oscillations, which are largely compensated, so that the evolution of the Lagrangian shear production is much smoother. The compensation points to the somewhat artificial split-up of the Lagrangian shear production into Eulerian- and Stokes shear production in such models. The model is not able to differentiate between the different energy sources. Remember, that the Stokes shear production is interpreted as a direct energy exchange between surface wave energy and $TKE$ \cite[]{Teixeira2002}.

\begin{figure}[hh]
  \centerline{\includegraphics[width=30pc,angle=0]{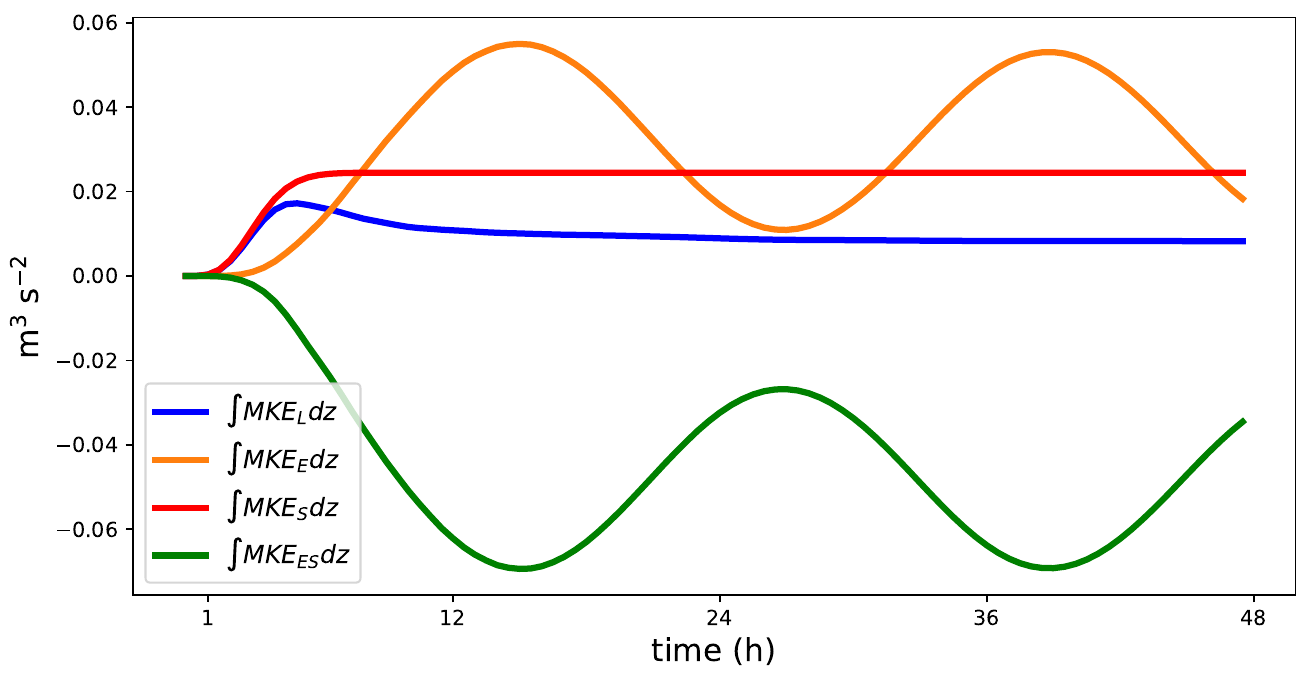}}
   \caption{Time evolution of vertically integrated mean kinetic energy compartments similar to figure 2 :Eulerian mean kinetic energy  $MKE_E$ (orange), Lagrangian mean kinetic energy  $MKE_L$ (blue),  the mixed  kinetic energy  $MKE_{ES}$ (green), and kinetic energy of the  Stokes drift $MKE_S$ (red). Note that  $MKE_S$ is solely based on the Stokes drift forcing which is identical to the laminar experiment. }
   \label{fig6}
\end{figure}

Similar to the energy loss to inertial oscillations, the energy loss to $TKE$ is only of minor importance for the energy budget of the surface waves on the here considered scales, but has strong consequences for 
the oceanic motions. On longer time and spatial scales, however, such an  energy loss might be an important contribution to swell dissipation  \cite[]{Ardhuin2006}.  

The vertically integrated mean kinetic energy compartments are given in Fig. 6. In contrast to the laminar case (Fig. \ref{fig3}),  $MKE_L$ drops to a value much lower than the kinetic energy in the Stokes drift  ($MKE_S$). The reduction in $MKE_L$ is caused by the energy loss to $TKE$. $MKE_E$ and $MKE_{ES}$ again exchange energy through the 
work done by the Coriolis-Stokes force. This exchange dominates the evolution of the $MKE_E$ budget. The loss to $TKE$ through the Eulerian shear production, i.e. term E1 in Eq.~(\ref{eq4}),
is not visible from this budget. $MKE_E$ is again difficult to interpret, as it contains the largest energy budget, but
most of this energy can not be  exchanged with other physically meaningful energy compartments.

\section{A framework for an energetically consistent coupling of a wave model to an ocean model}
 
 In this chapter we  discuss several issues related to a  large-scale general circulation ocean model coupled to a surface wave model. Special focus is given on consistent energy transfers. We concentrate on deep water waves and open ocean dynamics.
 
\subsection{Model equations}   

In the suggested framework, the ocean model integrates the Lagrangian velocity ${\mathbf u^L} = {\mathbf u}+{\mathbf u^S}$. The Stokes drift ${\mathbf u^S}$, and its evolution, will be provided by the wave model. In the open ocean, the horizontal gradients in the Stokes drift, as given by wave models, are governed by the atmospheric synoptic scales \cite[]{haney2015}, so that $\partial_x, \partial_y {\mathbf u^L} >> \partial_x, \partial_y {\mathbf u^S}$. The divergence in the Stokes drift is therefore expected to be small, and allows us to assume $w^S=0$, so that any horizontal divergence of the Stokes drift is compensated by $w$. 

Exploiting the assumptions for the open ocean in momentum equation (\ref{eq5}) and buoyancy equation (\ref {eq11}) , leads to the following equations to be used in large scale primitive equation ocean models:

\begin{equation}\label{eq22}
\partial_t {\mathbf u_h^L} + ({\mathbf u_h^L} \cdot \nabla) {\mathbf u^L} +  f{\mathbf z} \times  {\mathbf u^L}  = - \nabla_h p  + \nabla \mu_t  {\mathbf \nabla} ({\mathbf u^L} - {\mathbf u^S})  + \partial_t {\mathbf u_h^S} 
\end{equation}

\begin{equation}\label{eq23}
\partial_t b + ({\mathbf u_h^L} \cdot \nabla) b = \nabla \kappa_t  {\mathbf \nabla} b 
\end{equation}

Here, the subscript h denotes horizontal vector components, and  $\mu_t$ and $\kappa_t$ give turbulent viscosity and diffusivity, respectively. The Lagrangian velocity $ {\mathbf u^L}$ is the only prognostic velocity in the ocean model. Existing  numerical codes for advection, the Coriolis force, dissipation, and the buoyancy equation can be used. The prognostic model velocity is then re-interpreted as the Lagrangian velocity $u^L$, similar to the re-interpretation of the model velocity as residual mean velocity including the Quasi-Stokes velocity in the transformed residual mean theory of \cite{mcdougall2001}. Note, however, that dissipation should operate on ${\mathbf u} = {\mathbf u^L}-{\mathbf u^S}$. 

Although it is well known that $\nabla \cdot {\mathbf u^S} \ne 0$ \cite[]{McIntyre1988}, we ignore this divergence effect as it is usually quite small \cite[]{Vanneste2022}. The continuity equation is therefore given by $\nabla \cdot {\mathbf u^L}=0$.  

As a consequence of the open ocean assumption, the term $-{\mathbf u}^L \times  (\nabla \times \mathbf{u}^S)$ in Eq. (\ref{eq5}) can be neglected in the horizontal momentum equations (Eq.~(\ref{eq22})). Using scaling arguments, SFK16 show the possible importance of the term for the vertical momentum equation. SFK16 suggest to modify the hydrostatic balance in primitive equation models to a "wavy hydrostatic balance" of the form, $ \partial_z p - b = - u^L \partial_z u^S - v^L \partial_z v^S$. The necessity of including wave effects in the hydrostatic balance also depends on the resolved oceanic scales, and needs to be tested upon realization. Note, that using such a "wavy hydrostatic balance" would change the budget for $MKE_L$, and we recommend to use the "standard" hydrostatic balance, i.e. $\partial_z p = b$, for now.

\subsection{Energy and momentum fluxes}   

A wave model typically integrates a version of the wave energy balance equation. 
For deep water waves in the open ocean, it reads
\begin{equation}\label{eq24}
\partial_t F + \nabla \cdot ({\mathbf v_g} F)  = S_{in} + S_{nl} + S_{diss} 
\end{equation}
where $F(\omega, \theta)$ is the two-dimensional wave energy spectrum, which gives the energy distribution over angular frequency $\omega$ and propagation direction $\theta$. $\mathbf{v_g}$ is the group velocity. The r.h.s. of Eq.~(\ref{eq24}) gives the local source terms, which consists of wind input $S_{in}$, non-linear transfer $S_{nl}$, and dissipation due to wave breaking $S_{diss}$.

The source terms can be utilised to determine  the energy and momentum fluxes between wind, waves and ocean. The momentum flux ${\pmb \tau}_{in}$ and energy flux ${\Phi_{in}}$ from wind to the waves are given by \citep{Janssen2012}:
\begin{eqnarray}\label{eq25}
{\pmb \tau}_{in} &=& \rho_w g \int_0^{2 \pi} \int_0^{\infty} \frac{\mathbf k}{\omega}  \;  S_{in} \; d\omega d\theta 
\\
\label{eq25}
\Phi_{in} &=& \rho_w g \int_0^{2 \pi} \int_0^{\infty}  \;  S_{in} \; d\omega d\theta 
\end{eqnarray}
and the fluxes from the waves to the ocean column by:
\begin{eqnarray}\label{eq26}
{\pmb \tau}_{diss} &=& \rho_w g \int_0^{2 \pi} \int_0^{\infty} \frac{\mathbf k}{\omega}  \;  S_{diss} \; d\omega d\theta 
\\
\label{eq28}
\Phi_{diss} &=& \rho_w g \int_0^{2 \pi} \int_0^{\infty}  \;  S_{diss} \; d\omega d\theta 
\end{eqnarray}
Here, $\rho_w$ is the water density and ${\mathbf k}=(k_x,k_y)$ the wave number. Note, that the momentum fluxes are mostly determined
through the high frequency part of the wave spectrum, as they scale with the inverse of the phase velocity $g=\omega / k$.

The atmospheric or air-side stress is given by ${\pmb \tau}_{a} = \rho_a \mathbf{u_*^2}$, here $\rho_a$ is the air density  and $ \mathbf{u_*}$  the air friction velocity. The 
air-side stress is usually determined by a drag coefficient $c_d$ and the wind speed in 10m height. As the drag coefficient $c_d$ is dependent on the surface roughness, it should be
modified by the sea state, and the wave model can be used to determine the surface roughness (see e.g. \cite{Breivik2015}  for details).

The ocean side stress $\mathbf{\tau}_{oc}$ can be then considered as the atmospheric stress minus the residual momentum flux absorbed or released by the wave field

\begin{equation}\label{eq29}
{\pmb \tau}_{oc} = {\pmb \tau}_{a} - {\pmb \tau}_{in} - {\pmb \tau}_{diss}
\end{equation}

If wind increases over a calm ocean, the waves respond first by growing and ${\pmb \tau}_{in} > - {\pmb \tau}_{diss}$. As the waves mature, breaking intensifies, and so does the momentum transfer from the waves to the ocean ${\pmb \tau}_{diss}$. At some point  
during wave growth, ${\pmb \tau}_{diss}$ catch up with the 
momentum transfer from  atmosphere to waves ${\pmb \tau}_{in}$.
The wave field then is in equilibrium and ${\pmb \tau}_{in} = - {\pmb \tau}_{diss}$ \cite[]{Breivik2015}. At this point the ocean side stress equals the air-side stress and Eq. (\ref{eq29}) reduces to ${\pmb \tau}_{oc} = {\pmb \tau}_{a}$, which is the assumption made in classical bulk formulas, where ${\pmb \tau}_{oc}$ is a function of the wind speed in 10m height. However, most of the time such an equilibrium is a poor assumption \cite[]{hanley2010}. At some point the wind decreases, and the waves have a net momentum transfer into the ocean ${\pmb \tau}_{in} < - {\pmb \tau}_{diss}$. 

The prognostic frequency range in wave models has an upper limit $\omega_c$, above $\omega_c$ the wave spectrum is given by a widely accepted $\omega^{-5}$ power law. 
In the high-frequent diagnostic range, we assume that the wave field is always in equilibrium, so that ${\pmb \tau}_{oc} = {\pmb \tau}_{a}$ for $\omega > \omega_c$. The practical consequence is that Eq.~(\ref{eq29}) still holds,  if we integrate  ${\pmb \tau}_{in}$ and  ${\pmb \tau}_{diss}$ over the prognostic range.  A possible physical justification is provided by \cite{Chalikov1993}, they argue that the small waves are sheltered by the large waves and by-pass wave growth, thereby directly driving mean motions. Such an interpretation is of course an over-simplification of the problem. How exactly momentum fluxes enter the ocean is subject of active research, but beyond the scope of the present study.

Similar to the momentum fluxes, the energy flux into the ocean is given by \citep{Janssen2012}: 
\begin{equation}\label{eq30}
{\Phi}_{oc} = {\Phi}_{in} - \rho_w g \int_0^{2 \pi} \int_0^{\omega_c}  \;  (S_{in} + S_{diss}) \; d\omega d\theta 
\end{equation}
which can be also written as
\begin{equation}\label{eq31}
{\Phi}_{oc} = \rho_w g \int_0^{2 \pi} \int_{\omega_c}^{\infty}  \;  S_{in} \; d\omega d\theta \, -  \, {\Phi}_{diss} 
\end{equation}
i.e. the energy input consists of the direct energy gain from air in the diagnostic range and the dissipation of wave energy in the prognostic range, the latter is mainly the result of white capping. 
Note, the change in the limits of integration from Eq. (\ref{eq30}) to Eq. (\ref{eq31}).

${\Phi}_{oc}$ is the energy transfer from the surface waves available to drive oceanic mean motions and turbulence. The kinetic energy of the ocean model in the resolved velocities is governed by the tendency Eq.~(\ref{eq6}) for $MKE_L$.  
To allow for a consistent energy transfers to  sub-grid scales, we suggest using a second moment  closure, for example a
$k-\epsilon$ model. 
Such a closure integrates a TKE equation similar to   Eq.~(\ref{eq9}) for $TKE_L$, however the transfer terms are parameterized.
For simplicity we assume here a rigid-lid 
surface boundary condition, with $w^L=0$ and ${\pmb \tau}_{oc}= \overline{{\mathbf u}^{L \prime} w^{ \prime}}$ at  $z=0$.  The energy gain of the ocean column is then governed by the transport  terms T3 
and the forcing terms E7 in Eqs.~(\ref{eq6}) and  (\ref{eq9}). 
${\Phi}_{oc}$ should then match:

\begin{equation}\label{eq32}
{\Phi}_{oc} = \overline{{\mathbf u}^L} \cdot {\pmb \tau}_{oc} + \int \overline{{\mathbf u^L}} \cdot \partial_t  \overline{{\mathbf u^S} } \; dz  + \overline{{\mathbf u}^{L \prime} \cdot {\pmb \tau}^{\prime}_{oc}} + \int \overline{{\mathbf u^{L \prime}} \cdot \partial_t  {\mathbf u^{S \prime}} } \; dz + \epsilon_{break} 
\end{equation}
where $\epsilon_{break}$ denotes injection of $TKE$ by breaking waves. 

It is unclear, if either $\overline{{\mathbf u}^{L \prime} \cdot {\pmb \tau}^{\prime}_{oc}}$ or $\overline{{\mathbf u^{L \prime}} \cdot \partial_t  {\mathbf u^{S \prime}} }$ show any significant
correlation in the primed terms, so that they could be possibly neglected. However, we combine them together with the also unknown dissipation due to breaking of waves, so that 
\begin{equation}\label{eq33}
\Gamma_{break} = \overline{{\mathbf u}^{L \prime} \cdot {\pmb \tau}^{\prime}_{oc}} + \int \overline{{\mathbf u^{L \prime}} \cdot \partial_t  {\mathbf u^{S \prime}} } \; dz + \epsilon_{break} 
\end{equation}
$\Gamma_{break}$ can be determined in our coupled framework from Eq.~(\ref{eq32}), as the remaining terms can be obtained directly  from the ocean or the surface wave model.

\begin{figure}[hh]
  \centerline{\includegraphics[width=25pc,angle=0]{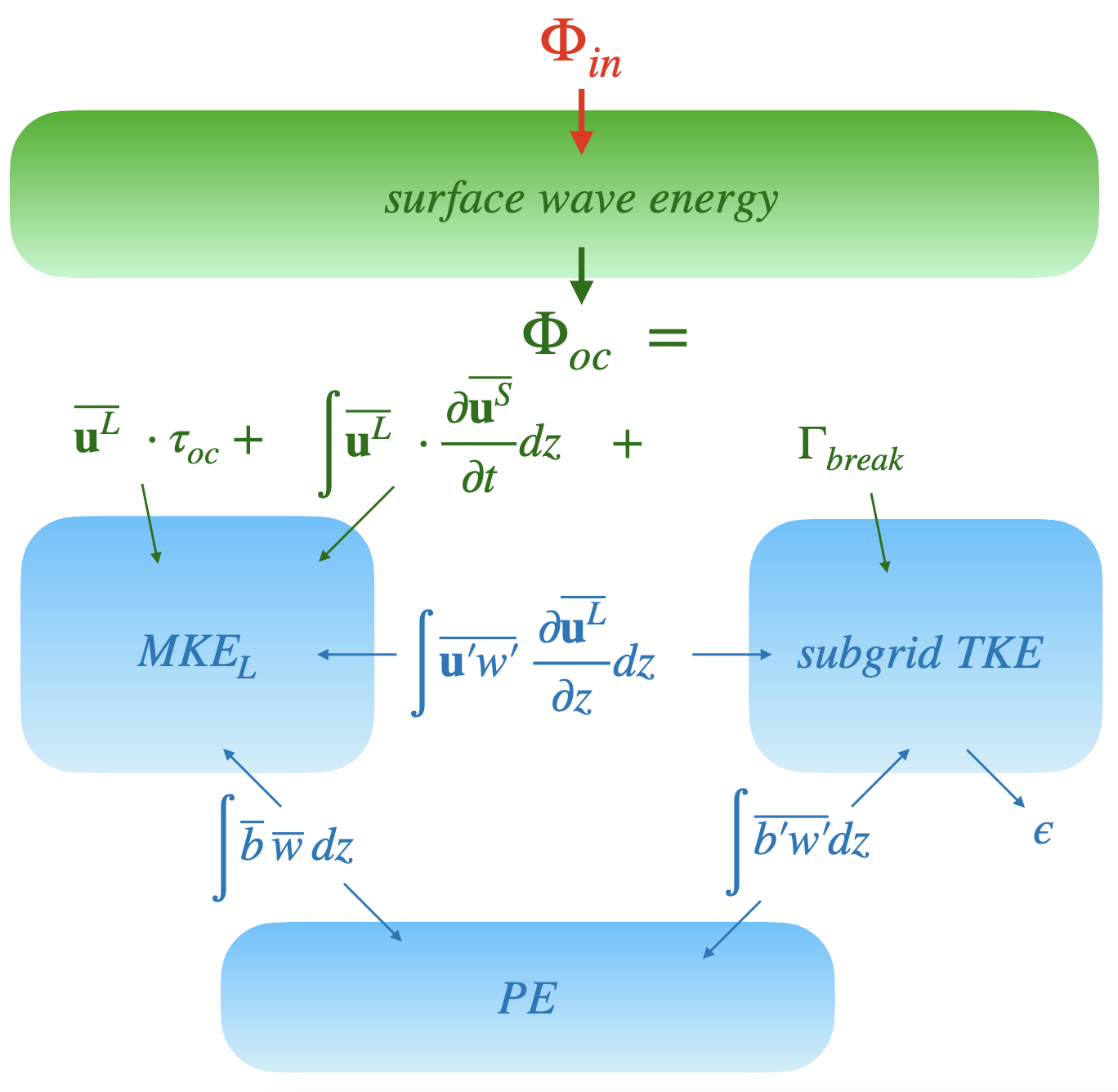}}
   \caption{Schematic of the energy exchanges between different compartments (coloured boxes). PE denotes potential energy and $\epsilon$ gives the integrated dissipation of
   subgrid $TKE$, i.e. the exchange with internal energy. Note, that the sign of the terms are associated with $MKE_L$ if possible.} 
   \label{fig7}
\end{figure}
Figure (\ref{fig7})  gives an overview of the different energy compartments and the involved energy transfers. The energy transfer from the wave model to the ocean is governed
by  Eq.~(\ref{eq32}).  $\Gamma_{break}$ goes directly into the subgrid $TKE$, whereas $\overline{{\mathbf u}^L} \cdot \overline{{\pmb \tau}_{oc}}$  and  $\int \overline{{\mathbf u^L}} \cdot \partial_t  \overline{{\mathbf u^S} } \; dz$ drive $MKE_L$, i.e. resolved oceanic motion. 

The Stokes production term $ -  \int  \overline{{\mathbf u^{L \prime}} w^{\prime}} \;\frac{\partial \overline{{\mathbf u^{S}}}}{ \partial z}\;dz$ could  be interpreted as a direct transfer from wave energy to subgrid $TKE$.  In our framework it is part of the Lagrangian shear production  term $ \int  \overline{{\mathbf u^{L \prime}} w^{\prime}} \;\frac{\partial \overline{{\mathbf u^{L}}}}{ \partial z}\;dz$,  which transfers energy between $MKE_L$ and  subgrid $TKE$. 
To be fully energetically consistent, the shear production term should
remove energy from the surface wave model and should be included in the wave dissipation term $\Phi_{diss}$ and therefore  $\Phi_{oc}$.
\cite{Ardhuin2010} in their recent update on wave dissipation parameterization, discussed the inclusion of the shear production term.
It was, however, neglected,  as its contribution is considered to be very small.

\subsection{Parameterizations}  

As mentioned above, we suggest to use a second moment closure for the subgrid $TKE$. \cite{Harcourt2013} and   \cite{Harcourt2015} provide such a closure using the Craik-Leibovich equations, i.e.
it involves a parameterization for the Stokes production term. The unknown Reynolds stresses are parameterized with two distinct  eddy viscosities $K_M$ and $K^S_M$, so that, e.g.  $\overline{u^{\prime} w^{\prime}} = - K_M \partial_z u - K^S_M \partial_z u^S$. The eddy viscosities are 
derived from stability functions  (see  \cite{Harcourt2013} for details). It should be however mentioned, that the distinct eddy viscosities lead to the same Reynolds stresses in  the Eulerian and in the Stokes production term, i.e. it does not resolve possible issues with overlapping wave and turbulence scales, as discussed in the previous section.
 
$\Gamma_{break}$ mainly consists of turbulence injected by breaking waves, the amount of energy is determined by eq. (\ref{eq32}). The energy can be injected as a flux 
boundary condition in the equations for the  second moment closure  following \cite{Burchard2001}.

\section{Summary and discussion}
 
Based on the wave-averaged Craik-Leibovich equations including a prescribed Stokes drift, we present a closed Lagrangian energy framework. The only energy exchange of the Lagrangian ocean flow
with surface waves is due to  changes in the Stokes drift forcing. Advantages compared to an 
Eulerian kinetic energy budget are that all energy transfer terms are well known and easy to interpret. 
In particular the work done by the "fictitious" Coriolis-Stokes force is absent in the Lagrangian energy budget.
Previous studies have suggested that  the work done by the  Coriolis-Stokes force is associated with an energy transfer from the surface waves to the Eulerian kinetic energy. We argue that this energy gain is an artefact of the split-up  of the Lagrangian kinetic energy 
into different compartments. The work done by the Coriolis-Stokes force is an exchange between the Eulerian kinetic energy and an energy compartment defined as $MKE_{ES} = \overline{{\mathbf u}}\cdot\overline{{\mathbf u}^S}$.  The individual compartments of the Lagrangian kinetic energy are physically difficult to interpret, but
the compensation between $MKE_{E}$ and $MKE_{ES}$ suggests that large parts of the  Eulerian mean kinetic energy are not available for a transfer to turbulent kinetic energy and finally for mixing. 
 
The ambiguity of the $MKE_{E}$ budget suggests that previous estimates of the energy input into the ocean by wind and waves should be interpreted with care.
Our Lagrangian framework suggests to ignore the work done by the Stokes-Coriolis force in such an estimate. The energy input into mean motions by the wind stress
is ${\mathbf u}^L \cdot \tau$  in our framework. At first glance this might be very different from previous estimates which used the Eulerian velocity or even the surface geostrophic velocity.  However, a Stokes drift is usually accompanied by an Eulerian anti-Stokes flow of similar order (at least vertically integrated), reducing the impact of the Stokes drift.

The compensation of Stokes drift by an anti-Stokes flow is very dependent on the impact of turbulence on both. Unfortunately, such an impact on surface waves and the associated
Stokes drift is largely unknown, hence ignored in most phase-averaged models . Nonetheless, the difference between Stokes drift shear and anti-Stokes flow shear plays an important role in the shear driven turbulence. The amount of compensation also modifies the surface Lagrangian velocity, and thus, the energy transfer from surface waves to mean motions. 

In the current phase-averaged ocean models, the same Reynolds stresses act on the Eulerian and the Stokes shear. Overlapping temporal and spatial scales of turbulence and surface waves suggest that this assumption might not be very realistic. It might be valid for Swell though. At present, there is no solution in phase-averaged models, and how turbulence effect surface waves is a future task for the phase resolving modelling community.

Using our Lagrangian framework, we suggest an energetically consistent coupling between a surface wave model and a large-scale ocean model. We recommend to use the Lagrangian velocity as prognostic velocity in the model as given by Eq. (\ref{eq22}). Other forms of the Craik-Leibovich equation are equally valid, but might demand more changes in an existing numerical code. The energy transfer from the waves to the ocean can then be split-up into energy which goes into mean motions and energy which goes into sub-grid turbulence. The transfer to mean motions consists of the work done by the surface stress, i.e.  $\overline{{\mathbf u}^L} \cdot \overline{{\pmb \tau}_{oc}}$,  and through temporal changes in the Stokes drift, $\int \overline{{\mathbf u^L}} \cdot \partial_t  \overline{{\mathbf u^S} } \; dz$. The latter term is expected to be much weaker than the work done by the surface stress, but is able to force, for example, strong surface wave driven inertial oscillations.
As the overall energy transfer from the surface waves model to the ocean $\Phi_{oc}$ can be estimated using Eq. (\ref{eq31}), the remainder energy transfer is related to wave breaking, which goes directly into subgrid turbulence.

The wave models usually allow to incorporate the sea surface velocity to be included in the second term of Eq. \ref{eq24}, which accounts for wave refraction by horizontal shear in $\mathbf{u}^L|_{z=0}$. The surface velocity could be also used in the wave model to compute the relative wind $\mathbf{u}^{atm}_{10} - \mathbf{u}^L|_{z=0}$, with $\mathbf{u}^{atm}_{10}$ being the atmospheric absolute wind at 10m. The relative wind rather than the absolute wind is often used in bulk formulations  for the atmospheric stress $\tau_a$. Note, however, that wave models seem very  sensitive to the bulk formulation (see \cite{Couvelard2020} for a detailed discussion).

We here describe a rather simple method to utilize the Lagrangian velocity in ocean models. More sophisticated methods are suggested, for example using the generalized Lagrangian mean \cite[]{Ardhuin2008} or vertically Lagrangian coordinates \cite[]{Aiki2012}. These phase-averaging methods follow the wave motions and allow for a concise treatment of the wavy surface. However, they are much more difficult to realize.

\clearpage
\acknowledgments
This paper is a contribution to the Collaborative Research Centre TRR 181
Energy Transfer in Atmosphere and Ocean funded by the Deutsche Forschungsgemeinschaft (DFG,
German Research Foundation) Projektnummer 274762653.

\datastatement
The numerical code, relevant model data, and scripts are available on Zenodo with the identifier
\url{https://doi.org/10.5281/zenodo.10043904}.


\bibliographystyle{ametsocV6}
\bibliography{literature}

\begin{thebibliography}{48}
\providecommand{\natexlab}[1]{#1}
\providecommand{\url}[1]{\texttt{#1}}
\renewcommand{\UrlFont}{\rmfamily}
\providecommand{\urlprefix}{URL }
\expandafter\ifx\csname urlstyle\endcsname\relax
  \providecommand{\doi}[1]{https://doi.org/\discretionary{}{}{}#1}\else
  \providecommand{\doi}{https://doi.org/\discretionary{}{}{}\begingroup
  \urlstyle{rm}\Url}\fi
\providecommand{\eprint}[2][]{\url{#2}}

\bibitem[{Aiki and Greatbatch(2012)Aiki, and Greatbatch}]{Aiki2012}
Aiki, H., and R.~J. Greatbatch, 2012: Thickness-weighted mean theory for the
  effect of surface gravity waves on mean flows in the upper ocean. \textit{J.\
  Phys.\ Oceanogr.}, \textbf{42~(5)}, 725--747.

\bibitem[{Ali et~al.(2019)Ali, Christensen, Breivik, Malila, Raj,
  an~E.~P.~Chassignet,, and Bakhoday-Paskyabi}]{Ali2019}
Ali, A., K.~H. Christensen, {\O}.~Breivik, M.~Malila, R.~P. Raj, L.~B.
  an~E.~P.~Chassignet, and M.~Bakhoday-Paskyabi, 2019: A comparison of
  {Langmuir} turbulence parameterizations and key wave effects in a numerical
  model of the {North Atlantic and Arctic Oceans}. \textit{Ocean Modelling},
  \textbf{137}, 76--97.

\bibitem[{Ardhuin and Jenkins(2006)Ardhuin, and Jenkins}]{Ardhuin2006}
Ardhuin, F., and A.~D. Jenkins, 2006: On the interaction of surface waves and
  upper ocean turbulence. \textit{J.\ Phys.\ Oceanogr.}, \textbf{36~(3)},
  551--557.

\bibitem[{Ardhuin et~al.(2008)Ardhuin, Rascle,, and Belibassakis}]{Ardhuin2008}
Ardhuin, F., N.~Rascle, and K.~A. Belibassakis, 2008: Explicit wave-averaged
  primitive equations using a generalized lagrangian mean. \textit{Ocean
  Modelling}, \textbf{20~(1)}, 35--60.

\bibitem[{Ardhuin et~al.(2010)}]{Ardhuin2010}
Ardhuin, F., and Coauthors, 2010: Semiempirical dissipation source functions
  for ocean waves. {Part I}: Definition, calibration, and validation.
  \textit{J.\ Phys.\ Oceanogr.}, \textbf{40~(9)}, 1917--1941.

\bibitem[{Belcher et~al.(2012)}]{belcher2012global}
Belcher, S.~E., and Coauthors, 2012: A global perspective on {{Langmuir}}
  turbulence in the ocean surface boundary layer. \textit{Geophys.\ Res.\
  Lett.}, \textbf{39~(18)}.

\bibitem[{Breivik et~al.(2015)Breivik, Mogensen, Bidlot, Balmaseda,, and
  Janssen}]{Breivik2015}
Breivik, {\O}., K.~Mogensen, J.-R. Bidlot, M.~A. Balmaseda, and P.~A. E.~M.
  Janssen, 2015: {Surface wave effects in the NEMO ocean model: Forced and
  coupled experiments}. \textit{J.\ Geophys.\ Res., Oceans}, \textbf{120~(4)},
  2973--2992.

\bibitem[{Brostr{\"o}m et~al.(2014)Brostr{\"o}m, Christensen, Drivdal,, and
  Weber}]{Brostrom2014}
Brostr{\"o}m, G., K.~H. Christensen, M.~Drivdal, and J.~E.~H. Weber, 2014: Note
  on {Coriolis-Stokes} force and energy. \textit{Ocean Dynamics}, \textbf{64},
  1039--1045.

\bibitem[{Burchard(2001)}]{Burchard2001}
Burchard, H., 2001: Simulating the wave-enhanced layer under breaking surface
  waves with two-equation turbulence models. \textit{J.\ Phys.\ Oceanogr.},
  \textbf{31~(11)}, 3133--3145.

\bibitem[{Chalikov and Belevich(1993)Chalikov, and Belevich}]{Chalikov1993}
Chalikov, D., and M.~Y. Belevich, 1993: One-dimensional theory of the wave
  boundary layer. \textit{Boundary-Layer Meteorology}, \textbf{63}, 65--96.

\bibitem[{Couvelard et~al.(2020)Couvelard, Lemari{\'e}, Samson, Redelsperger,
  Ardhuin, Benshila,, and Madec}]{Couvelard2020}
Couvelard, X., F.~Lemari{\'e}, G.~Samson, J.-L. Redelsperger, F.~Ardhuin,
  R.~Benshila, and G.~Madec, 2020: Development of a two-way-coupled ocean--wave
  model: assessment on a global nemo (v3. 6)--ww3 (v6. 02) coupled
  configuration. \textit{Geosci. Model Dev.}, \textbf{13~(7)}, 3067--3090.

\bibitem[{Craik and Leibovich(1976)Craik, and Leibovich}]{Craik76}
Craik, A. D.~D., and S.~Leibovich, 1976: A rational model for {Langmuir}
  circulations. \textit{J.\ Fluid\ Mech.}, \textbf{73~(03)}, 401--426.

\bibitem[{Czeschel and Eden(2019)Czeschel, and Eden}]{Czeschel2019}
Czeschel, L., and C.~Eden, 2019: Internal wave radiation through surface mixed
  layer turbulence. \textit{J.\ Phys.\ Oceanogr.}, \textbf{49~(7)}, 1827--1844.

\bibitem[{Ducros et~al.(1996)Ducros, Comte,, and Lesieur}]{Ducros1996}
Ducros, F., P.~Comte, and M.~Lesieur, 1996: Large-eddy simulation of transition
  to turbulence in a boundary layer developing spatially over a flat plate.
  \textit{J.\ Fluid\ Mech.}, \textbf{326}, 1--36.

\bibitem[{Eden et~al.(2014)Eden, Czeschel,, and Olbers}]{Eden2014}
Eden, C., L.~Czeschel, and D.~Olbers, 2014: Toward energetically consistent
  ocean models. \textit{J.\ Phys.\ Oceanogr.}, \textbf{44~(12)}, 3160--3184.

\bibitem[{Fan and Griffies(2014)Fan, and Griffies}]{Fan2014}
Fan, Y., and S.~M. Griffies, 2014: Impacts of parameterized {Langmuir}
  turbulence and nonbreaking wave mixing in global climate simulations.
  \textit{J.\ Climate}, \textbf{27~(12)}, 4752--4775.

\bibitem[{Haney et~al.(2015)Haney, Fox-Kemper, Julien,, and Webb}]{haney2015}
Haney, S., B.~Fox-Kemper, K.~Julien, and A.~Webb, 2015: Symmetric and
  geostrophic instabilities in the wave-forced ocean mixed layer. \textit{J.\
  Phys.\ Oceanogr.}, \textbf{45~(12)}, 3033--3056.

\bibitem[{Hanley et~al.(2010)Hanley, Belcher,, and Sullivan}]{hanley2010}
Hanley, K.~E., S.~E. Belcher, and P.~P. Sullivan, 2010: A global climatology of
  wind--wave interaction. \textit{J.\ Phys.\ Oceanogr.}, \textbf{40~(6)},
  1263--1282.

\bibitem[{Harcourt(2013)}]{Harcourt2013}
Harcourt, R.~R., 2013: A second-moment closure model of {Langmuir} turbulence.
  \textit{J.\ Phys.\ Oceanogr.}, \textbf{43~(4)}, 673--697.

\bibitem[{Harcourt(2015)}]{Harcourt2015}
Harcourt, R.~R., 2015: An improved second-moment closure model of {Langmuir}
  turbulence. \textit{J.\ Phys.\ Oceanogr.}, \textbf{45~(1)}, 84--103.

\bibitem[{Hasselmann(1970)}]{hasselmann1970wave}
Hasselmann, K., 1970: Wave-driven inertial oscillations. \textit{Geophy. and
  Astrophys. Fluid Dyn.}, \textbf{1~(3-4)}, 463--502.

\bibitem[{Holm(1996)}]{Holm1996}
Holm, D.~D., 1996: The ideal {Craik-Leibovich} equations. \textit{Physica D:
  Nonlinear Phenomena}, \textbf{98~(2-4)}, 415--441.

\bibitem[{Huang(1979)}]{Huang1979}
Huang, N.~E., 1979: On surface drift currents in the ocean. \textit{J.\ Fluid\
  Mech.}, \textbf{91~(1)}, 191--208.

\bibitem[{Janssen(2012)}]{Janssen2012}
Janssen, P.~A., 2012: Ocean wave effects on the daily cycle in sst. \textit{J.\
  Geophys.\ Res., Oceans}, \textbf{117~(C11)}.

\bibitem[{Komen et~al.(1996)Komen, Cavaleri, Donelan, Hasselmann, Hasselmann,,
  and Janssen}]{komen1996}
Komen, G.~J., L.~Cavaleri, M.~Donelan, K.~Hasselmann, S.~Hasselmann, and
  P.~Janssen, 1996: \textit{Dynamics and modelling of ocean waves}. Cambridge
  University Press.

\bibitem[{Leibovich(1980)}]{Leibovich1980}
Leibovich, S., 1980: On wave-current interaction theories of {Langmuir}
  circulations. \textit{J.\ Fluid\ Mech.}, \textbf{99~(4)}, 715--724.

\bibitem[{Li et~al.(2016)Li, Webb, Fox-Kemper, Craig, Danabasoglu, Large,, and
  Vertenstein}]{Li2016}
Li, Q., A.~Webb, B.~Fox-Kemper, A.~Craig, G.~Danabasoglu, W.~G. Large, and
  M.~Vertenstein, 2016: {Langmuir mixing effects on global climate: WAVEWATCH
  III in CESM}. \textit{Ocean Modelling}, \textbf{103}, 145--160.

\bibitem[{Liu et~al.(2009)Liu, Wu,, and Guan}]{Liu2009wind}
Liu, B., K.~Wu, and C.~Guan, 2009: Wind energy input to the ekman-stokes layer:
  reply to comment by jeff a. polton. \textit{Journal of Oceanography},
  \textbf{65}, 669--673.

\bibitem[{McDougall and McIntosh(2001)McDougall, and McIntosh}]{mcdougall2001}
McDougall, T.~J., and P.~C. McIntosh, 2001: The temporal-residual-mean
  velocity. part ii: Isopycnal interpretation and the tracer and momentum
  equations. \textit{J.\ Phys.\ Oceanogr.}, \textbf{31~(5)}, 1222--1246.

\bibitem[{McIntyre(1988)}]{McIntyre1988}
McIntyre, M., 1988: A note on the divergence effect and the {Lagrangian-mean}
  surface elevation in periodic water waves. \textit{J.\ Fluid\ Mech.},
  \textbf{189}, 235--242.

\bibitem[{McWilliams et~al.(2014)McWilliams, Huckle, Liang,, and
  Sullivan}]{mcwilliams2014}
McWilliams, J.~C., E.~Huckle, J.~Liang, and P.~P. Sullivan, 2014: Langmuir
  turbulence in swell. \textit{J.\ Phys.\ Oceanogr.}, \textbf{44~(3)},
  870--890.

\bibitem[{McWilliams et~al.(1997)McWilliams, Sullivan,, and
  Moeng}]{McWilliams1997}
McWilliams, J.~C., P.~P. Sullivan, and C.-H. Moeng, 1997: Langmuir turbulence
  in the ocean. \textit{J.\ Fluid\ Mech.}, \textbf{334}, 1--30.

\bibitem[{Olbers et~al.(2012)Olbers, Willebrand,, and Eden}]{Olbers2012}
Olbers, D., J.~Willebrand, and C.~Eden, 2012: \textit{Ocean dynamics}. Springer
  Science \& Business Media.

\bibitem[{Pollard(1970)}]{Pollard1970}
Pollard, R.~T., 1970: Surface waves with rotation: An exact solution.
  \textit{J.\ Geophys.\ Res.}, \textbf{75~(30)}, 5895--5898.

\bibitem[{Polton(2009)}]{Polton2009}
Polton, J.~A., 2009: A wave averaged energy equation: Comment on “global
  estimates of wind energy input to subinertial motions in the {Ekman-Stokes}
  layer” by bin liu, kejian wu and changlong guan. \textit{Journal of
  Oceanography}, \textbf{65}, 665--668.

\bibitem[{Polton et~al.(2005)Polton, Lewis,, and Belcher}]{Polton2005}
Polton, J.~A., D.~M. Lewis, and S.~E. Belcher, 2005: The role of wave-induced
  coriolis--stokes forcing on the wind-driven mixed layer. \textit{J.\ Phys.\
  Oceanogr.}, \textbf{35~(4)}, 444--457.

\bibitem[{Sayol et~al.(2016)Sayol, Orfila,, and Oey}]{Sayol2016}
Sayol, J.~M., A.~Orfila, and L.-Y. Oey, 2016: Wind induced energy--momentum
  distribution along the {Ekman--Stokes} layer. application to the {Western
  Mediterranean Sea climate}. \textit{Deep Sea Research Part I: Oceanographic
  Research Papers}, \textbf{111}, 34--49.

\bibitem[{Skyllingstad and Denbo(1995)Skyllingstad, and
  Denbo}]{Skyllingstad1995}
Skyllingstad, E.~D., and D.~W. Denbo, 1995: An ocean large-eddy simulation of
  {Langmuir} circulations and convection in the surface mixed layer.
  \textit{J.\ Geophys.\ Res.}, \textbf{100~(C5)}, 8501--8522.

\bibitem[{Stokes(1847)}]{Stokes1847}
Stokes, G.~G., 1847: On the theory of oscillatory waves. \textit{Trans. Cam.
  Philos. Soc.}, \textbf{8}, 441--455.

\bibitem[{Sun et~al.(2022)}]{Sun2022}
Sun, R.~A., and Coauthors, 2022: {Waves in SKRIPS: WaveWatch III coupling
  implementation and a case study of cyclone Mekunu}. \textit{EGUsphere}.

\bibitem[{Suzuki and Fox-Kemper(2016)Suzuki, and Fox-Kemper}]{Suzuki2016}
Suzuki, N., and B.~Fox-Kemper, 2016: Understanding {Stokes} forces in the
  wave-averaged equations. \textit{J.\ Geophys.\ Res.}, \textbf{121~(5)},
  3579--3596.

\bibitem[{Teixeira and Belcher(2002)Teixeira, and Belcher}]{Teixeira2002}
Teixeira, M. A.~C., and S.~E. Belcher, 2002: On the distortion of turbulence by
  a progressive surface wave. \textit{J.\ Fluid\ Mech.}, \textbf{458},
  229--267.

\bibitem[{Ursell and Deacon(1950)Ursell, and Deacon}]{Ursell1950}
Ursell, F., and G.~E.~R. Deacon, 1950: {On the theoretical form of ocean swell.
  On a rotating earth}. \textit{Geophysical Journal International}, \textbf{6},
  1--8.

\bibitem[{Vanneste and Young(2022)Vanneste, and Young}]{Vanneste2022}
Vanneste, J., and W.~R. Young, 2022: Stokes drift and its discontents.
  \textit{Philos.\ Trans.\ Roy.\ Soc.\ London}, \textbf{380~(2225)},
  20210\,032.

\bibitem[{Wagner et~al.(2021)Wagner, Chini, Ramadhan, Gallet,, and
  Ferrari}]{Wagner2021}
Wagner, G.~L., G.~P. Chini, A.~Ramadhan, B.~Gallet, and R.~Ferrari, 2021:
  Near-inertial waves and turbulence driven by the growth of swell. \textit{J.\
  Phys.\ Oceanogr.}, \textbf{51~(5)}, 1337--1351.

\bibitem[{WAVEWATCH~III(2016)}]{wavewatch2016}
WAVEWATCH~III, R., 2016: Development group {(WW3DG)}: User manual and system
  documentation of {WAVEWATCH III R} version 5.16. \textit{Tech. Note 329,
  NOAA/NWS/NCEP/MMAB, College Park, MD, USA}.

\bibitem[{Weber et~al.(2015)Weber, Drivdal, Christensen,, and
  Brostr{\"o}m}]{Weber2015}
Weber, J. E.~H., M.~Drivdal, K.~H. Christensen, and G.~Brostr{\"o}m, 2015: Some
  aspects of the {Coriolis-Stokes} forcing in the oceanic momentum and energy
  budgets. \textit{Journal of Geophysical Research: Oceans}, \textbf{120~(8)},
  5589--5596.

\bibitem[{Zhang et~al.(2019)Zhang, Song, Wu,, and Shi}]{Zhang2019}
Zhang, Y., Z.~Song, K.~Wu, and Y.~Shi, 2019: Influences of random surface waves
  on the estimates of wind energy input to the {Ekman} layer in the {Antarctic}
  circumpolar current region. \textit{Journal of Geophysical Research: Oceans},
  \textbf{124~(5)}, 3393--3410.

\end{thebibliography}

\end{document}